\documentclass[a4paper,12pt]{article}
\usepackage[dvips]{graphics}
\usepackage{calc,pstcol,multido,pst-node,amsgen,amsmath,amsfonts,amssymb,bm}
\usepackage{rotating}
\usepackage[nolists,nomarkers,tablesfirst]{endfloat}
\usepackage{lscape,tikz}
\usetikzlibrary{positioning,shapes,arrows.meta}
\usepackage{booktabs} 

\newcommand{\ci}{\perp\!\!\!\perp}

\newcommand{\tm}{ t \text{-} }
\newcommand{\sm}{ s \text{-} }

\newcommand{\zerop}{ 0 \text{+} }
\newcommand{\tdeltam}{ (t + \delta) \text{-} }

\begin{document}
\parindent=0pt
\parskip=5pt

\begin{center}
{\LARGE \bf   Simulating data from marginal structural models for a survival time outcome}
\end{center}

\begin{center}
  {\large \bf   Shaun R.\ Seaman$^1$ and Ruth H.\ Keogh$^2$} \\
  \vspace{.1cm}
  23/12/2023
\end{center}


$^1$ MRC Biostatistics Unit, University of Cambridge, East Forvie Building, University Forvie Site, Robinson Way, Cambridge, CB2 0SR, UK. \\
shaun.seaman@mrc-bsu.cam.ac.uk

$^2$ Department of Medical Statistics, London School of Hygiene and Tropical Medicine, Keppel Street, London, WC1E 7HT, UK. \\
Ruth.Keogh@lshtm.ac.uk

\vspace{0.1cm}

\section*{Abstract}

Marginal structural models (MSMs) are often used to estimate causal effects of treatments on survival time outcomes from observational data when time-dependent confounding may be present.
They can be fitted using, e.g., inverse probability of treatment weighting (IPTW).
It is important to evaluate the performance of statistical methods in different scenarios, and simulation studies are a key tool for such evaluations.
In such simulation studies, it is common to generate data in such a way that the model of interest is correctly specified, but this is not always straightforward when the model of interest is for potential outcomes, as is an MSM.
Methods have been proposed for simulating from MSMs for a survival outcome, but these methods impose restrictions on the data-generating mechanism.
Here we propose a method that overcomes these restrictions.
The MSM can be a marginal structural logistic model for a discrete survival time or a Cox or additive hazards MSM for a continuous survival time.
The hazard of the potential survival time can be conditional on baseline covariates, and the treatment variable can be discrete or continuous.
We illustrate the use of the proposed simulation algorithm by carrying out a brief simulation study.
This study compares the coverage of confidence intervals calculated in two different ways for causal effect estimates obtained by fitting an MSM via IPTW.

\vspace{0.5cm}

Key words: bootstrap, causal inference, compatible models, congenial models, continuous-time marginal structural model, sandwich estimator, simulation studies, survival analysis, time-dependent confounding.

\newpage

\section{Introduction}

In many longitudinal observational studies, a time-varying treatment and set of time-varying covariates are observed for each individual at a number of time points (`visits').
A frequent goal with such studies is to estimate the causal effect of treatment on some outcome of interest, e.g.\ survival time.
Such estimation is often complicated by time-dependent confounding: an individual's treatment at a particular visit depends on that individual's covariate history and may affect values of covariates at future visits\cite{Daniel2013}.
Estimating causal effects in this setting requires specific statistical methods, of which marginal structural models (MSMs) is the most popular\cite{Clare2019}.
A MSM is a model for the potential (or `counterfactual') outcome that would arise if treatment were assigned according to a particular rule.
MSMs are frequently fitted by inverse probability of treatment weighting (IPTW).
A commonly-used class of MSMs for a survival outcome is Cox MSMs\cite{Hernan2000,Sterne2005,Clare2019}.
Here we focus on MSMs for a survival outcome, including Cox MSMs and their discrete-time analogues, and consider static treatment rules, i.e.\ where treatment allocation is pre-specified, rather than depending on information collected during the study (see Section~\ref{sect:discussion} for dynamic treatment regimes).

It is important to evaluate performance of new (and existing) statistical methods in different scenarios.
Simulation studies are a key tool for such evaluations, allowing the assessment of, e.g., bias, coverage of confidence intervals (CI), relative efficiency compared to other methods, and robustness to violation of assumptions\cite{Morris2019,Friedrich2023}.
It is common in such studies to generate data in such a way that the model of interest is correctly specified.
It transpires this is not always simple when the model of interest is a model for potential outcomes\cite{Evans2024}, which is the case for MSMs.

Several authors have proposed methods for simulating from MSMs for a survival outcome.
Methods for simulating from Cox MSMs were proposed by Xiao et al.\ (2010)\cite{Xiao2010} and Young et al.\ (2010)\cite{Young2010}.
Young and Tchetgen Tchetgen (2014)\cite{Young2014} and Havercroft and Didelez (2012)\cite{Havercroft2012} proposed algorithms for simulating from, respectively, a discrete-time Cox MSM and a discrete-time marginal structural logistic model.
By finely discretising time, these two algorithms can be used to simulate a continuous failure time from a Cox MSM.
Keogh et al.\ (2021)\cite{Keogh2021} described how to simulate from an additive-hazards MSM.

These existing methods impose restrictions on the data-generating mechanism.
Xiao et al.\ (2010) rely on the failure rate being very low, do not allow for (possibly unobserved) common causes of the time-dependent confounders and survival process, and (as noted by~\cite{Young2014}) the hazard in the implied MSM depends on all past treatments.
Havercroft and Didelez (2012) assume the dependence of the survival process on the time-dependent confounder process arises entirely through the effect of the latter on the treatment process and through a single shared latent variable.
This means, for example, that the effect of confounders measured at time 1 on the hazard of failure at time 5, say, is as strong as their effect on the hazard at time 2, rather than being able to diminish over time.
Young et al.\ (2010) seek to generate data in such a way that a Cox MSM, a structural nested accelerated failure time model and a structural nested failure time model are all correctly specified, so that performance of these three modelling approaches can be compared.
Their method is closely related to that of Havercroft and Didelez (2012) but requires that the potential treatment-free survival time be exponentially distributed and the hazard in the Cox MSM depend only on the most recent treatment; also treatment must be binary.
Young and Tchetgen Tchetgen (2014) aim to simulate data in such a way that three other models are correctly specified: a Cox MSM, a parametric model for the treatment given the past, and a parametric model for the hazard given the past.
This allows the comparison of two methods for fitting Cox MSMs: IPTW and g-computation.
Their data-generating mechanism assumes there are no baseline confounders and only one time-dependent confounder, that the hazard at time $t$ depends only on the treatment at the two most recent times and the time-dependent confounder at the most recent time, and that the time-dependent confounder at time $t$ depends on the history of treatment and confounder only through treatment at the most recent time.
In addition, none of these four data-simulation methods explicitly allows for a MSM that conditions the hazard on baseline covariates.
Keogh et al.'s (2021)\cite{Keogh2021} proposal for simulating from an additive-hazards MSM does not impose such restrictions, but also does not enable the user easily to specify the values of the parameters of the MSM directly.

In this article, we propose an algorithm for simulating from a MSM for a survival time outcome that overcomes these restrictions.
This MSM can be, for example, a marginal structural logistic model for a discrete survival time, or a Cox or additive hazards MSM for a continuous survival time.
The hazard can be conditional on baseline covariates, and the treatment variable can be discrete or continuous.
In developing our algorithm, we have broadly followed the general and powerful approach proposed by Evans and Didelez (2024)\cite{Evans2024} for simulating from causal models.
They sketched how this approach might be used to generate from a MSM for survival data but provided few details and then only for the case of one binary confounder with a strong Markov property (see our Appendix~\ref{appendix:YTT} for details).

In Section~\ref{sect:notation} we introduce notation, define the MSM of interest and present the causal directed acyclic graph (DAG) assumed to apply.
Section~\ref{sect:K1} describes our proposed algorithm when the survival time is discrete.
It requires the specification of a `risk score' function and the correlation parameter of a Gaussian copula.
The risk score is a function of the history of the confounders, and the copula expresses the association between this risk score and the potential survival outcome.
Application of the algorithm requires the cumulative distribution function (CDF) of the risk score to be known, which it will not be in general.
So, in Section~\ref{sect:manyAX}, we extend our algorithm to estimate this CDF at the same time as simulating the data.
Section~\ref{sect:cns.time} details how the algorithm can be adapted for a continuous failure time; this enables simulation from a Cox or additive hazards MSM.
A brief simulation study is presented in Section~\ref{sect:simulationstudy}, to illustrate the methods described in this article.
{\tt R} code for implementing this study is provided in Supplementary Materials, along with Appendices~\ref{appendix:YTT}--\ref{appendix:discrete}.
Section~\ref{sect:discussion} contains a discussion.

\section{Set-up and notation}

\label{sect:notation}

Consider a study in which $n$ individuals are observed at regular visits up to the earlier of their failure time and censoring time.
Visit times, assumed to be the same for all individuals, are $0, 1, \ldots K$, and the administrative censoring time is $K+1$.
For convenience, we shall sometimes refer to the adminstrative censoring time as `visit $K+1$'.
We denote random variables with capital letters and their values with lower-case letters.

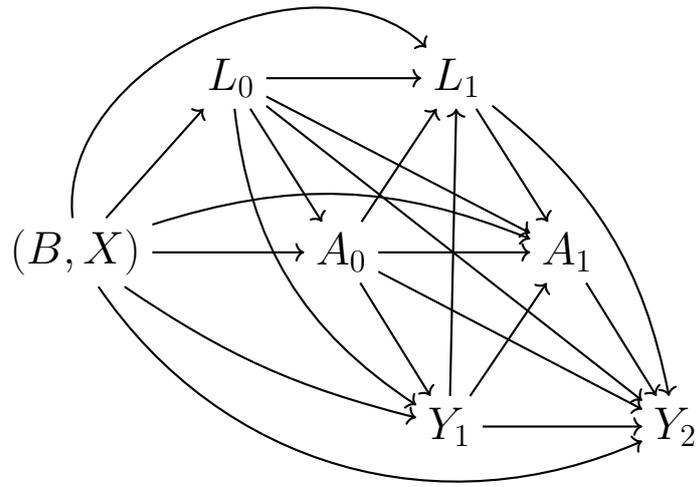
\begin{figure}
\begin{center}
\begin{tikzpicture}[auto, node distance=2cm, thick, node/.style={font=\sffamily\Large}]
  \node[node] (BX) {$(B, X)$};
  \node[node] (A0) [right = 2cm of BX] {$A_0$};
  \node[node] (A1) [right = 2cm of A0] {$A_1$};
  \node[node] (L0) [above left = 1.5cm and 0.5cm of A0] {$L_0$};
  \node[node] (L1) [above left = 1.5cm and 0.5cm of A1] {$L_1$};
  \node[node] (Y1) [below right = 1.5cm and 0.5cm of A0] {$Y_1$};
  \node[node] (Y2) [below right = 1.5cm and 0.5cm of A1] {$Y_2$};
  \path[every node/.style={font=\sffamily\small}]
  (BX) edge[->] node [right] {} (L0)
  (BX) edge[->] node [right] {} (A0)
  (BX) edge[->,bend right=10] node [right] {} (Y1)
  (BX) edge[->,bend left=70] node [right] {} (L1)
  (BX) edge[->,bend right=40] node [right] {} (Y2)
  (BX) edge[->,bend left=20] node [right] {} (A1)
  (A0) edge[->] node [right] {} (A1)
  (A0) edge[->] node [right] {} (L1)
  (A0) edge[->] node [right] {} (Y1)
  (A0) edge[->] node [right] {} (Y2)
  (A1) edge[->] node [right] {} (Y2)
  (L0) edge[->] node [right] {} (A0)
  (L0) edge[->] node [right] {} (A1)
  (L0) edge[->] node [right] {} (L1)
  (L0) edge[->,bend right=25] node [right] {} (Y1)
  (L0) edge[->] node [right] {} (Y2)
  (L1) edge[->] node [right] {} (A1)
  (L1) edge[->,bend left=20] node [right] {} (Y2)
  (Y1) edge[->] node [right] {} (A1)
  (Y1) edge[->] node [right] {} (Y2)
  (Y1) edge[->] node [right] {} (L1);
\end{tikzpicture}
\end{center}
\caption{Assumed causal directed acyclic graph (DAG).  For simplicity, this is shown for $K=1$.}
\label{fig:causalDAG}
\end{figure}

Let $A_k$ denote the treatment received by an individual at visit $k$ ($k=0, \ldots, K$).
This could be discrete or continuous.
Let $T$ denote the individual's failure time and $Y_k=I(T \geq k)$ be an indicator of survival to visit $k$ ($k=1, \ldots, K+1$), with $Y_0 = 1$.
Let $X$ and $B$ denote two distinct vectors of baseline covariates for an individual.
Variables $X$ are baseline confounders and/or treatment effect modifiers that will be conditioned on in the MSM presented below (equation~(\ref{eq:msm})).
Variables $B$ are not conditioned on in the MSM, and can include any or all of: baseline confounders, common causes of the $L_k$'s and $Y_k$'s variables that are not confounders, and instrumental variables.
Either of $X$ and $B$ could be empty.
Let $L_k$ denote time-varying confounders for an individual at visit $k$ ($k=0, \ldots, K$).
Let $\bar{A}_k = (A_0, \ldots, A_k)$ and $\bar{L}_k = (L_0, \ldots, L_k)$ denote the histories of treatment and time-dependent confounders up to visit $k$; and $\bar{A}_{-1} = \bar{L}_{-1} = \emptyset$.
We assume the causal DAG shown in Figure~\ref{fig:causalDAG}.

Variables with a superscript $\bar{a}_k$ are potential variables under an intervention that sets treatments $\bar{A}_k$ at the first $k+1$ visits equal to $\bar{a}_k$.
For example, $Y_{k+1}^{\bar{a}_k} = 1$ if the individual would survive to visit $k+1$ if their treatments at visits $0, \ldots, k$ were set to $\bar{a}_k$, and $Y_{k+1}^{\bar{a}_k} = 0$ if the individual would fail.
We make the usual consistency assumption that $Y_k = Y_k^{\bar{A}_{k-1}}$ and $L_k = L_k^{\bar{A}_{k-1}}$.
Note that this together with the causal DAG implies sequential exchangeability, i.e.\
$
(Y_{k+1}^{\bar{A}_{k-1} \; \underline{a}_k}, \ldots, Y_{K+1}^{\bar{A}_{k-1} \; \underline{a}_k}) \ci A_k \mid X, B, \bar{L}_k, \bar{A}_{k-1}, Y_k=1
$
for all $\underline{a}_k = (a_k, \ldots, a_K)$.

We begin with MSMs for discrete-time hazard of failure, postponing consideration of MSMs for continuous-time hazard, e.g.\ Cox MSMs, until Section~\ref{sect:cns.time}.
Consider the following MSM for the hazard at visit $k$ ($k=1, \ldots, K+1$).
\begin{equation}
P( Y_k^{\bar{a}_{k-1}} = 0 \mid X, Y_{k-1}^{\bar{a}_{k-2}} = 1 ) = g_k (\bar{a}_{k-1}, X; \beta),
\label{eq:msm}
\end{equation}
where $g_k(.)$ is a known function with parameters $\beta$.
For example, in a marginal structural logistic model\cite{Robins2000},
$
g_k (\bar{a}_{k-1}, X; \beta) = \mbox{expit} \left\{ \sum_{j=1}^J \beta_{kj} \; q_{kj} (\bar{a}_{k-1}, X)) \right\},
$
where $q_{k1}, \ldots, q_{kJ}$ are known functions of $\bar{a}_{k-1}$ and $X$, e.g.\ $g_k (\bar{a}_{k-1}, X; \beta) = \mbox{expit} (\beta_{k0} + \beta_{k1}^\top X + \beta_{k2} a_{k-1})$.

We shall describe a method for simulating data on $(X, B, L_0, A_0, Y_1, \ldots, L_K, A_K, Y_{K+1})$ for a single individual such that equation~(\ref{eq:msm}) is satisfied for a pre-specified choice of functions $g_k (\bar{a}_{k-1}, X; \beta)$ and values of $\beta_{kj}$ ($k=1, \ldots, K+1$).
By applying this method $n$ times, data for $n$ individuals can be simulated.

\section{Algorithm for simulating data}

\label{sect:K1}

For simplicity, consider the case of $K=1$ (Figure~\ref{fig:causalDAG}); we extend the proposed method to $K \geq 1$ in Section~\ref{sect:generalK}.

The joint distribution of $(X, B, L_0, A_0, Y_1, L_1, A_1, Y_2)$ can be factorised as
\begin{eqnarray}
  p(X, B, L_0, A_0, Y_1, L_1, A_1, Y_2)
  & = &
        p(X) \; P(B, L_0 \mid X) \; P(A_0 \mid X, B, L_0)
  \nonumber \\
  && \times
     p(Y_1 \mid X, B, L_0, A_0) \; p(L_1 \mid X, B, L_0, A_0, Y_1)
     \nonumber \\
  && \times
     p(A_1 \mid X, B, L_0, A_0, Y_1, L_1)
          \nonumber \\
  && \times
p(Y_2 \mid X, B, L_0, A_0, Y_1, L_1, A_1).
     \label{eq:factorisation}
\end{eqnarray}
For each individual in the sample, we shall generate the variables in the order shown by equation~(\ref{eq:factorisation}).
If the first three distributions on the right-hand side of equation~(\ref{eq:factorisation}) are specified, $(X, B, L_0, A_0)$ can be sampled from them.
If the fifth and sixth distributions are also specified, we shall be able to sample $(L_1, A_1)$ from them once we have sampled $Y_1$.

This leaves the task of sampling $Y_1$ given $(X, B, L_0, A_0)$ and $Y_2$ given $(X, B, L_0, A_0, Y_1, L_1, A_1)$.
By the consistency assumption, some potential survival indicators $Y_1^{a_0}$ and $Y_2^{\bar{a}_1}$ are related to the observed indicators $Y_1$ and $Y_2$: specifically, $Y_1 = Y_1^{A_0}$ and $Y_2 = Y_2^{\bar{A}_1}$.
So, if we want (as we do) the MSM of equation~(\ref{eq:msm}) to be correctly specified for the chosen functions $g_k (\bar{a}_{k-1}, X; \beta)$ and values of $\beta_{kj}$ ($k=1,2$), then we need to choose the distributions $p(Y_1 \mid X, B, L_0, A_0)$ and $p(Y_2 \mid X, B, L_0, A_0, Y_1, L_1, A_1)$ --- or equivalently, $p(Y_1^{A_0} \mid X, B, L_0, A_0)$ and $p(Y_2^{\bar{A}_1} \mid X, B, L_0, A_0, Y_1, L_1, A_1)$ --- carefully.
For example, sampling $Y_1$ from a simple logistic regression model with covariates $X$, $B$, $L_0$ and $A_0$ will typically not lead to equation~(\ref{eq:msm}) being satisfied.
We now describe suitable methods for sampling $Y_1$ and $Y_2$.

\subsection{Sampling $Y_1$ given $(X, B, L_0, A_0)$}

\label{sect:generate.Y1}

As just noted, the task of sampling $Y_1$ given $(X, B, L_0, A_0)$ is the same as that of sampling $Y_1^{A_0}$ given $(X, B, L_0, A_0)$.
We shall consider the general task of sampling $Y_1^{a_0}$ given $(X, B, L_0, A_0)$ for any value of $a_0$.
If we can do this, then we can sample $Y_1^{A_0}$ specifically and then set $Y$ equal to it.
Let $p(Y_1^{a_0} \mid X, B, L_0, A_0)$ denote a conditional distribution that satisfies the sequential exchangeability assumption, i.e.\ $p(Y_1^{a_0} \mid X, B, L_0, A_0) = p(Y_1^{a_0} \mid X, B, L_0)$.
We call this a `potential sampling distribution' (for $Y_1^{a_0}$).
For equation~(\ref{eq:msm}) to hold when $k=1$, $p(Y_1^{a_0} \mid X, B, L_0)$ needs to satisfy
\begin{equation}
\int P(Y_1^{a_0} = 0 \mid X, B, L_0) \; p(B, L_0 \mid X) \; dB \; dL_0
=
P( Y_1^{a_0} = 0 \mid X) = g_1 (a_0, X; \beta).
\label{eq:integrate}
\end{equation}
Many distributions of $Y_1^{a_0}$ given $(X, B, L_0)$ would satisfy this constraint, although it is not immediately obvious how to find one that does.

Our solution to the problem of finding (or more precisely, sampling from) a potential sampling distribution that satisfies equation~(\ref{eq:integrate}) involves two components: a `risk score' function and a bivariate Gaussian copula\cite{Schepsmeier2014}.
The first will allow us to reduce the dependence of $Y_1^{a_0}$ on $B$ and $L_0$ to dependence on a scalar function of those variables, which we call a `risk score'.
The second will describe the association between this risk score and a latent continuous variable (denoted $U_{Y_1^{a_0}}$ in what follows) that determines $Y_1^{a_0}$.
The user of our algorithm will specify the risk score function and the correlation parameter in the copula.

Let $h_0^{a_0} (x, b, l_0)$ be a scalar continuous function of $(x, b, l_0)$ that the user of our algorithm should specify.
We call this the `risk score' function for visit 0, because it will rank individuals with the same value of $X$ but different values of $(B, L_0)$ according to their potential hazards $P(Y_1^{a_0} = 0 \mid X, B, L_0)$.
We call the scalar continuous random variable $H_0^{a_0} = h_0^{a_0} (X, B, L_0)$ the `risk score' (for visit 0) for an individual (under the intervention that sets $A_0 = a_0$).
Let $F_{H_0^{a_0}} (h \mid x)$ denote the conditional CDF of $H_0^{a_0}$ given $X=x$, and let $U_{H_0^{a_0}} = F_{H_0^{a_0}} (H_0^{a_0} \mid X)$ be the random variable obtained by plugging the individual's variables $H_0^{a_0}$ and $X$ into this CDF.
We call $U_{H_0^{a_0}}$ the individual's `risk quantile'.
Notice that (by a general property of CDFs) $U_{H_0^{a_0}} \mid X \sim \mbox{Uniform} (0,1)$.

Let $\Phi(.)$ denote the CDF of the standard normal distribution and $\rho_0$ ($-1 < \rho_0 \leq 0$) be some constant that the user of our algorithm will choose (in principle, $\rho_0$ could be a function of $a_0$, but for simplicity, we shall assume it is not).
We shall show that the potential sampling distribution
\begin{equation}
  P(Y_1^{a_0} = 0 \mid X, B, L_0) =
  \Phi \left(
    \frac{ \Phi^{-1} \{ g_1 (a_0, X; \beta) \}  - \rho_0 \Phi^{-1} (U_{H_0^{a_0}}) }
    { \sqrt{1 - \rho_0^2} }
    \right)
    \label{eq:copula.conditional}
\end{equation}
satisfies equation~(\ref{eq:msm}) for $k=1$ and that the algorithm we present below samples $Y_1^{a_0}$ from this distribution.
\begin{figure}
\begin{center}
  \scalebox{0.8} {
    \includegraphics{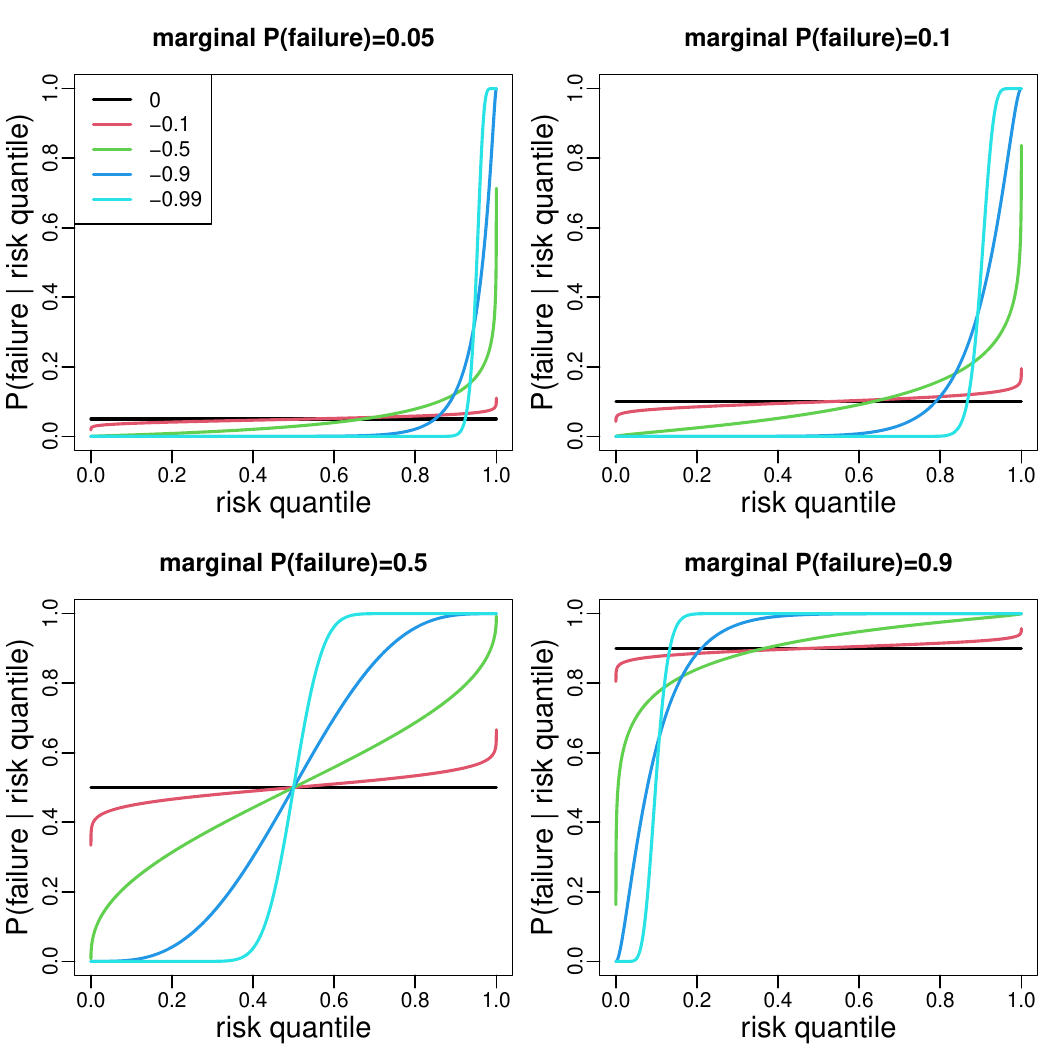}
  }
\end{center}
\caption{Probability $P(Y_1^{a_0} = 0 \mid X, B, L_0)$ of failure as a function of the risk quantile $U_{H_0^{a_0}}$ for five different values of $\rho_0$ and four different marginal probabilities of failure $g_1 (a_0, X; \beta)$.}
\label{fig:copulaplots}
\end{figure}

Note that $U_{H_0^{a_0}}$, and hence $P(Y_1^{a_0} = 0 \mid X, B, L_0)$, does not change if the risk score function $h_0^{a_0} (x, b, l_0)$ is replaced by $\nu \{ h_0^{a_0} (x, b, l_0) \}$, where $\nu(.)$ is any monotonically increasing function.
What matters therefore is only the {\it ranking} of individuals that $h_0^{a_0} (x, b, l_0)$ provides.
This and the fact that the right-hand side of equation~(\ref{eq:copula.conditional}) is an increasing function of $U_{H_0^{a_0}}$, which itself is uniformly distributed, justifies our naming $U_{H_0^{a_0}}$ the `risk quantile'.
Parameter $\rho_0$ describes the strength of dependence of $P(Y_1^{a_0} = 0 \mid X, B, L_0)$ on $U_{H_0^{a_0}}$.
At one extreme, there is no dependence when $\rho_0=0$; at the other, $Y_1^{a_0} \approx I \{ U_{H_0^{a_0}} < g_1 (a_0, X; \beta) \}$ when $\rho \approx -1$ (see Figure~\ref{fig:copulaplots}).

An example of a possible risk score function is $h_0^{a_0} (x, b, l_0) = 1^\top l_0$, i.e.\ the sum of the elements of the vector $l_0$.
This risk score function means that of two individuals who have the same value of $X$ but different values of $(B, L_0)$, the individual with the higher value of $1^\top L_0$ has the greater potential hazard (when $A_0$ is set to $a_0$ by an intervention).
The reason why $h_0^{a_0} (x, b, l_0)$ is written with a superscript $a_0$ and as a function of $x$ is that user can, if desired, specify that the dependence of $P(Y_1^{a_0} = 0 \mid X, B, L_0)$ on $B$ and $L_0$ is different for different values of $a_0$ and/or $X$.
For example, if $A_0$ is binary and $h_0^0 (x, b, l_0) = l_0$ and $h_0^1 (x, b, l_0) = 0$, then $L_0$ affects an individual's potential hazard only if treatment is set to zero.

Our algorithm involves the following steps.
First, calculate $H_0^{a_0}$, then $U_{H_0^{a_0}} = F_{H_0^{a_0}} (H_0^{a_0} \mid X)$ and $Z_{H_0^{a_0}} = \Phi^{-1} (U_{H_0^{a_0}})$.
Then generate $Z_{Y_1^{a_0}} \sim \mbox{Normal} (\rho_0 Z_{H_0^{a_0}}, 1 - \rho_0^2)$ and calculate $U_{Y_1^{a_0}} = \Phi (Z_{Y_1^{a_0}})$.
Finally, set
\begin{equation}
  Y_1^{a_0} = 0 \mbox{ if } U_{Y_1^{a_0}} < g_1 (a_0, X; \beta) \mbox{ and } Y_1^{a_0} = 1 \mbox{ otherwise}.
  \label{eq:Y.from.U}
\end{equation}
We now show that this algorithm generates $Y_1^{a_0}$ from equation~(\ref{eq:copula.conditional}) and satisfies equation~(\ref{eq:msm}) for $k=1$.
First, notice that $Z_{H_0^{a_0}} \mid X \sim \mbox{Normal} (0,1)$, and hence that $(Z_{H_0^{a_0}}, Z_{Y_1^{a_0}})$ has a bivariate normal distribution given $X$ with means 0, variances 1, and correlation $\rho_0$.
This implies that
$
U_{Y_1^{a_0}} \mid X \sim \mbox{Uniform} (0,1).
$
%
It follows from this and expression~(\ref{eq:Y.from.U}) that equation~(\ref{eq:msm}) is satisfied for $k=1$.
Also, equation~(\ref{eq:copula.conditional}) immediately follows from
\begin{eqnarray*}
  &&
  P(Y_1^{a_0} = 0 \mid X, B, L_0)
  = 
  P( U_{Y_1^{a_0}} < g_1 (a_0, X; \beta) \mid X, B, L_0)
  \\
  && \hspace{0.5cm} = 
  P \left(
  \frac{ \Phi^{-1} ( U_{Y_1^{a_0}} ) - \rho_0 Z_{H_0^{a_0}} }
       { \sqrt{1 - \rho_0^2} }
       <
       \frac{ \Phi^{-1} \{ g_1 (a_0, X; \beta) \} - \rho_0 Z_{H_0^{a_0}} }
         { \sqrt{1 - \rho_0^2} }
         \mid X, B, L_0
         \right).
\end{eqnarray*}

Readers familiar with copulas will recognise that we have used a Gaussian copula to describe the conditional association between $U_{H^{a_0}}$ and $U_{Y_1^{a_0}}$ given $X$, and sampled $U_{Y_1^{a_0}}$ from its conditional distribution given $(U_{H^{a_0}}, X)$ implied by this copula.
A non-Gaussian copula could be used instead, which would give a different function of $U_{H_0^{a_0}}$ on the right-hand side of equation~(\ref{eq:copula.conditional}).
However, whereas the Gaussian copula causes $P(Y_1^{a_0} = 0 \mid X, B, L_0)$ to be an increasing function of $U_{H_0^{a_0}}$, this is not true of all copulas.
Thus, although $U_{H_0^{a_0}}$ would still represent the function of $(B, L_0)$ through which $Y_1^{a_0}$ depends on $B$ and $L_0$ (c.f.\ the linear predictor in a logistic regression model), it might not be interpretable as a risk quantile.

\subsection{Sampling $Y_2$ given $(X, B, L_0, A_0, Y_1, L_1, A_1)$}

\label{sect:sample.Y2}

Suppose we have generated $Y_1 = Y_1^{A_0}$ using the method of Section~\ref{sect:generate.Y1}.
If $Y_1 = 0$, we stop: the individual has failed before visit 1.
Otherwise, generate $L_1$ and then $A_1$ from the fifth and sixth distributions on the right-hand side of equation~(\ref{eq:factorisation}).
The final step is to sample $Y_2$ given $(X, B, L_0, A_0, L_1, A_1)$ and $Y_1=1$.
The procedure we now propose for doing this is analogous to that proposed in Section~\ref{sect:generate.Y1} to generate $Y_1$.
This time, we need to specify: i) a risk score function that ranks individuals with the same value of $X$ and $Y_1^{a_0}=1$ but different values of $(B, L_0, L_1^{a_0})$ by their hazards, and ii) the correlation parameter $\rho_1$ of a Gaussian copula for the conditional association (given $X$ and $Y_1^{a_0}=1$) between the risk quantile and a latent continuous variable $U_{Y_2^{\bar{a}_1}}$ that determines $Y_2^{\bar{a}_1}$.

We shall consider the task of generating $Y_2^{A_0 a_1}$ given the previously generated variables $(X, B, L_0, A_0, Y_1, L_1, A_1)$ for any value of $a_1$.
If we can do this, then we can generate $Y_2^{\bar{A}_1}$ specifically and set $Y_2$ equal to it.
The causal DAG and consistency assumption together imply
%
$
P(Y_2^{\bar{a}_1} = 0 \mid X, B, L_0, \bar{A}_1 = \bar{a}_1, Y_1=1, L_1)
  =
  P(Y_2^{\bar{a}_1} = 0 \mid X, B, L_0, Y_1^{a_0}=1, L_1^{a_0})
$
%
%
for any $\bar{a}_1$ (see Appendix~\ref{appendix:proof} for proof).
When, as here, $a_0$ is the previously generated value of $A_0$, the consistency assumption implies that $L_1^{a_0} = L_1$ and $Y_1^{a_0} = Y_1$ are known (i.e.\ have been previously generated).

Let $h_1^{\bar{a}_1} = h_1^{\bar{a}_1} (x, b, l_0, l_1)$ be a scalar continuous function of $(x, b, l_0, l_1)$, specified by the user.
This is the `risk score' function for visit 1, and $H_1^{\bar{a}_1} = h_1^{\bar{a}_1} (X, B, L_0, L_1^{a_0})$ is the individual's `risk score' (for visit 1).
Let $F_{H_1^{\bar{a}_1}} (h \mid x, Y_1^{a_0}=1)$ denote the CDF of $H_1^{\bar{a}_1}$ given $X=x$ and $Y_1^{a_0}=1$, and call $U_{H_1^{\bar{a}_1}} = F_{H_1^{\bar{a}_1}} (H_1^{\bar{a}_1} \mid X, Y_1^{a_0} = 1)$ the individual's `risk quantile' (for visit 1).

Let $Z_{H_1^{\bar{a}_1}} = \Phi^{-1} (U_{H_1^{\bar{a}_1}})$ and generate $Z_{Y_2^{\bar{a}_1}} \sim \mbox{Normal} (\rho_1 Z_{H_1^{\bar{a}_1}}, 1 - \rho_1^2)$.
Calculate $U_{Y_2^{\bar{a}_1}} = \Phi (Z_{Y_2^{\bar{a}_1}})$ and set $Y_2^{\bar{a}_1} = 0$ if $U_{Y_2^{\bar{a}_1}} < g_2 (\bar{a}_1, X; \beta)$ and $Y_2^{\bar{a}_1} = 1$ otherwise.
This ensures that equation~(\ref{eq:msm}) is satisfied for $k=2$ and $Y_2^{\bar{a}_1}$ is generated from
\[
P(Y_2^{\bar{a}_1} = 0 \mid X, B, L_0, Y_1^{a_0} = 1, L_1^{a_0}) = 
  \Phi \left(
    \frac{ \Phi^{-1} \{ g_2 (\bar{a}_1, X; \beta) \}  - \rho_1 \Phi^{-1} (U_{H_1^{\bar{a}_1}}) }
    { \sqrt{1 - \rho_1^2} }
    \right).
    \]

Examples of possible $h_1^{\bar{a}_1} (x, b, l_0, l_1)$ are $1^\top l_1$ and $1^\top (l_1 - l_0)$.
The latter implies that the potential hazard of failure (when $\bar{A}_1$ is set to $\bar{a}_1$ by an intervention) depends on the most recent change in the values of the time-dependent confounders.

\subsection{Generalisation to more than two visits}

\label{sect:generalK}

The method just described for $K=1$ generalises to any $K \geq 1$.
First, specify distributions $p(X)$ and $p(B \mid X)$, and $p(L_k \mid \bar{A}_{k-1}, \bar{L}_{k-1}, Y_k = 1)$ and $p(A_k \mid \bar{A}_{k-1}, \bar{L}_k, Y_k = 1)$ for each $k=0, \ldots, K$.
Also for each $k=0, \ldots, K$, specify a scalar continuous risk score function $h_k^{\bar{a}_k} (x, b, \bar{l}_k)$ and correlation parameter $-1 < \rho_k \leq 0$ of a Gaussian copula for the conditional association between risk quantile $U_{H_k^{\bar{a}_k}}$ and latent variable $U_{Y_{k+1}^{\bar{a}_k}}$ (both defined below) given $X$ and $Y_k^{\bar{a}_{k-1}} = 1$.
The risk score function will rank individuals with the same value of $X$ and $Y_k^{\bar{a}_{k-1}}=1$ but different values of $(B, \bar{L}_k^{\bar{a}_{k-1}})$.
Now,
\begin{enumerate}
\item
  Sample $X$ from $p(X)$ and then $B$ from $p(B \mid X)$.
  Set $k=0$.
\item
  Sample $L_k$ from $p(L_k \mid X, B, \bar{L}_{k-1}, \bar{A}_{k-1}, Y_k=1)$.
\item
  Sample $A_k$ from $p(A_k \mid X, B, \bar{L}_k, \bar{A}_{k-1}, Y_k=1)$ and call the result $a_k$.
\item
  Calculate $H_k^{\bar{a}_k} = h_k^{\bar{a}_k} (X, B, \bar{L}_k)$.
\item
  Calculate $U_{H_k^{\bar{a}_k}} = F_{H_k^{\bar{a}_k}} (H_k^{\bar{a}_k} \mid X, Y_k^{\bar{a}_{k-1}}=1)$ and $Z_{H_k^{\bar{a}_k}} = \Phi^{-1} (U_{H_k^{\bar{a}_k}})$.
\item
  Sample $Z_{Y_{k+1}^{\bar{a}_k}} \sim \mbox{Normal} (\rho_k Z_{H_k^{\bar{a}_k}}, 1 - \rho_k^2)$ and calculate $U_{Y_{k+1}^{\bar{a}_k}} = \Phi ( Z_{Y_{k+1}^{\bar{a}_k}} )$.
\item
  If $U_{Y_{k+1}^{\bar{a}_k}} < g_{k+1} (\bar{a}_k, X; \beta)$, set $Y_{k+1}=0$.
  Otherwise set $Y_{k+1}=1$.
\item
  If $Y_{k+1} = 1$ and $k<K$, let $k=k+1$ and return to step 2.
\end{enumerate}

Note that we set $H_k^{\bar{a}_k} = h_k^{\bar{a}_k} (X, B, \bar{L}_k)$ at step 4 because the consistency assumption means that $\bar{L}_k^{\bar{a}_k} = \bar{L}_k$ for this value of $\bar{a}_k$, and we are implicitly setting $Y_{k+1} = Y_{k+1}^{\bar{a}_k}$ at step 7 for the same reason.
Note also that this algorithm does not require the rejection sampling that Evans and Didelez (2024) use in general\cite{Seaman2024}.

If censoring can occur before time $K+1$, the individual's censoring time can be sampled after applying steps 1--8, using whatever censoring mechanism is desired.
Alternatively, sampling of a censoring time can be embedded within these steps.

\section{Extended algorithm for estimating CDF of risk score and simulating data}

\label{sect:manyAX}

For a limited range of data-generating mechanisms, the CDF $F_{H_k^{\bar{a}_k}} (h \mid x, Y_k^{\bar{a}_{k-1}} = 1)$ of the risk score $H_k^{\bar{a}_k}$ given $X=x$ and $Y_k^{\bar{a}_{k-1}} = 1$ is known (see final paragraph of Appendix~A).
In general, however, it is not, which poses an obstacle to implementing the algorithm of Section~\ref{sect:K1}.
When the number of possible values of $(X, \bar{A}_K)$ is small (e.g.\ because treatment is binary and does not cease once started and there are no baseline covariates in the MSM),  $F_{H_k^{\bar{a}_k}} (h \mid x, Y_k^{\bar{a}_{k-1}} = 1)$ can be estimated separately for each of these values using the procedure described in Appendix~\ref{appendix:knownF}.

More generally, one can use the following extended algorithm, which involves simultaneously generating data and estimating the CDFs.
This is done by generating data not only for a single sampled individual, but also for a moderately large number (here we use 4999) of matching individuals at the same time.
Each of these matches has the same values of $X$ and $\bar{A}_K$ as the sampled individual, but its values of $B$, $L_k$ and $Y_k$ are randomly generated (in the same way as for the sampled individual), and so will usually differ from the sampled individual's values.
These 5000 individuals, i.e.\ the one sampled individual and that individual's 4999 matches, are used to estimate their values of $F_{H_k^{\bar{a}_k}} (H_k^{\bar{a}_k} \mid X, Y_k^{\bar{a}_{k-1}} = 1)$ immediately before sampling their $Y_{k+1}$ values.
Once the data for the sampled individual have been generated, the data on the 4999 matches are discarded.

The algorithm for generating data for a {\it single} sampled individual is as follows.
In this algorithm, the single sampled individual and the 4999 matches of this individual are indexed by $j$, with $j=1$ being the sampled individual and $j=2, \ldots, m$ being the matches, where $m = 5000$.
A random variable not mentioned so far, $I_j$ ($j=2, \ldots, m$), appears in the algorithm.
We shall make other comments about the algorithm after stating it.

\begin{enumerate}
\item
  Sample $X_1$ from $p(X)$.
  Let $X_j = X_1$ and $I_j = j$ for $j=2, \ldots, m$.
  Set $k=0$.
\item
  For $j=1, \ldots, m$, sample $B_j$ from $p(B_j \mid X_j)$.
\item
  For $j=1, \ldots, m$, sample $L_{kj}$ from $p(L_{kj} \mid X_j, B_j, \bar{L}_{k-1,j}, \bar{A}_{k-1,1} = \bar{a}_{k-1}, Y_{kj}=1)$.
\item
  Sample $A_{k1}$ from $p(A_{k1} \mid X_1, B_1, \bar{L}_{k1}, \bar{A}_{k-1,1} = \bar{a}_{k-1}, Y_{k1}=1)$ and call the result $a_k$.
\item
  For $j=1, \ldots, m$, calculate $H_{kj} = H_k^{\bar{a}_k} (X_j, B_j, \bar{L}_{kj})$.
  \item
  For $j=1, \ldots, m$, Let $R_{kj}$ denote the rank of $H_{kj}$ among the set $\{H_{k1}, \ldots, H_{km} \}$, sample $W_{kj} \sim \mbox{Uniform} (0,1)$, and calculate $U_{H_k^{\bar{a}_k}, j} = (R_{kj} - W_{kj}) / m$.
\item
  For $j=1, \ldots, m$, calculate $Z_{H_k^{\bar{a}_k}, j} = \Phi^{-1} (U_{H_k^{\bar{a}_k}, j})$.
\item
  For $j=1, \ldots, m$,   Sample $Z_{Y_{k+1}^{\bar{a}_k}, j} \sim \mbox{Normal} (\rho_k Z_{H_k^{\bar{a}_k}, j}, 1 - \rho_k^2)$ and calculate $U_{Y_{k+1}^{\bar{a}_k}, j} = \Phi ( Z_{Y_{k+1}^{\bar{a}_k}, j} )$.
\item
  For $j=1, \ldots, m$, set $Y_{k+1,j}=0$ if $U_{Y_{k+1}^{\bar{a}_k},j} < g_{k+1} (\bar{a}_k, X_j; \beta)$ and set $Y_{k+1,j}=1$ otherwise.
\item
  If $Y_{k+1, 1} = 0$ or $k=K$, stop.
\item
  For $j=2, \ldots, m$, if $Y_{k+1,j} = 0$, randomly choose $2 \leq j^* \leq m$ such that $Y_{k+1,j^*} = 1$, and set $I_j = I_{j^*}$, $B_j = B_{j^*}$, $\bar{L}_{kj} = \bar{L}_{kj^*}$ and $Y_{k+1,j} = 1$.
\item
    Let $k=k+1$ and return to step 3.
\end{enumerate}

Readers wishing to understand this algorithm should note that $H_{k1}, \ldots, H_{km}$ (in step 5) constitute a random sample from the distribution of $H_k^{\bar{a}_{k-1}}$ given $X$ and $Y_k=1$, and hence $R_{kj} / m$ (in step 6) is an estimate of $F_{H_k^{\bar{a}_k}} (H_{kj}^{\bar{a}_k} \mid X_j, Y_{kj}^{\bar{a}_{k-1}} = 1)$.
The reason for subtracting $W_{kj}$ from $R_{kj}$ (in step 6) is to ensure that the conditional distribution of $U_{H_k j}^{\bar{a}_k}$ given $X_j$ and $Y_{kj}=1$ is Uniform$(0,1)$.

The purpose of step 11 is to replace each match that fails at time $k+1$ with a copy of a randomly chosen match that has not yet failed, thus maintaining the number of matches at $m-1 = 4999$.
The variables $I_2, \ldots, I_m$ keep track of which matches are copies of which other matches.
At first, $I_j = j$ for all $j$ (see step 1), but each time a match, say match $j$, fails and is replaced by a copy of another match, say match $j^*$, $I_j$ is changed to $j^*$ (step 11).

A potential problem with this algorithm is the possibility that eventually the number of distinct values of $\{ I_2, \ldots, I_m \}$ could be small, indicating that few of the original 4999 matches remain.
This could cause the algorithm to perform poorly, because it would mean, for example, that the matches only had a small number of distinct values of $(B, L_1)$.
This is most likely to happen when $P( Y_k^{\bar{a}_{k-1}} = 0 \mid X)$ for the sampled individual's value of $X$ is close to one.

To address this problem, we replace step 12 above with a new step 12 and add an additional step 13.
The purpose of step 13 is to discard the existing matches and restart with a much larger number (we use $m=100000$) of matches whenever few (e.g.\ less than 10\%) of the original 4999 matches remain at step $k^*+1$.
The data generated for the sampled individual ($j=1$) up to visit $k^*$ and their indicator, $Y_{k^*+1,1}$, of survival to visit $k^*+1$ are retained.
The new steps 12 and 13 are:

\begin{enumerate}
  \setcounter{enumi}{11}
\item
  If the number of unique values in the set $\{ I_2, \ldots, I_m \}$ is greater than some threshold, e.g.\ $m/10$,
  then let $k=k+1$ and return to step 3.
\item
  Let $k^*$ denote the current value of $k$, i.e.\ the visit reached just before entering this step.
  Set $k=0$, $m=100000$ and $X_j = X_1$ $(j=2, \ldots, m)$,
  \begin{enumerate}
  \item
    Repeat step 2 for $j=2, \ldots, m$ (i.e.\ for the matches only).
  \item
    Repeat step 3, but only include $j=1$ if $k>k^*$.
  \item
    Repeat step 4 if $k>k^*$.
  \item
    Repeat steps 5--8.
  \item
    Repeat step 9, but only include $j=1$ if $k>k^*$.
  \item
    Repeat steps 10 and 11.
  \item
    Let $k=k+1$ and return to step 12b.
  \end{enumerate}
\end{enumerate}


In Appendix~\ref{appendix:empirical.study}, we demonstrate this algorithm does indeed generate data compatible with the chosen MSM, by simulating data on one million individuals, fitting the MSM to these data and verifying that the parameter estimates obtained are very close to the true parameter values.
We also present (Appendix~\ref{appendix:sensitivity}) an investigation of sensitivity of results to the choice of $m$, where we find that $m=1000$ in steps 1--11 and $m=20000$ in step 13 is probably adequate.

\section{Continuous failure time}

\label{sect:cns.time}

We now extend the earlier algorithms to simulate data for a continuous-time failure model, e.g.\ Cox or additive hazards MSM.

The first step is to specify how $\lambda^{\bar{a}_K} (t \mid X=x)$, the continuous-time hazard at time $t$ of the potential failure time $T^{\bar{a}_K}$ when we intervene to set $\bar{A}_K = \bar{a}_K$, depends on $x$ and $\bar{a}_K$.
In a Cox MSM, this has the form
\[
  \lambda^{\bar{a}_K} (t \mid X=x; \; \beta) =
  \lambda^{\bar{0}_K} (t \mid X=0)
\times \exp \left\{ \sum_{j=1}^J \beta_{kj} q_{kj} (\bar{a}_k, x) \right\}
\hspace{0.6cm}
(k < t \leq k+1)
\]
where $\lambda^{\bar{0}_K} (t \mid X=0)$ is the potential hazard for an individual with $X=0$ when all treatments are set equal to zero, $q_{kj} (\bar{a}_k, x)$ are known functions of $\bar{a}_k$ and $x$, and $\beta_{kj}$ are parameters.
In an additive hazards MSM, $\lambda^{\bar{a}_K} (t \mid X=x; \; \beta) = \lambda^{\bar{0}_K} (t \mid X=0) + \sum_{j=1}^J \beta_{kj} q_{kj} (\bar{a}_k, x)$, or the same formula but with $\beta_{kj}$ replaced by $\beta_j(t)$.
When fitting a MSM, the analyst specifies the functions $q_{kj} (\bar{a}_k, x)$, e.g.\ $q_{k1} = x$ and $q_{k2} = a_k$, treats the $\beta$ parameters as unknown quantities to be estimated, and places no restriction on $\lambda^{\bar{0}_K} (t \mid X=0)$.
To simulate data for a MSM, however, we need to specify all of $q_{kj} (\bar{a}_k, x)$, $\beta$ and $\lambda^{\bar{0}_K} (t \mid X=0)$.
Henceforth, assume these quantities have been specified.

By definition, $Y_k^{\bar{a}_K} = I(T^{\bar{a}_K} > k)$ ($k=1, \ldots, K+1$).
The conditional CDF of $T^{\bar{a}_k}$ given $X$ and $T^{\bar{a}_k} \geq k$ is, when $k < t \leq k+1$,
\begin{equation}
F_{T^{\bar{a}_k}} (t \mid X=x, T^{\bar{a}_k} \geq k)
=
1 - \exp \left \{ - \int_k^t \lambda^{\bar{a}_K} (s \mid X=x; \; \beta) \; ds \right\}.
\label{eq:FT.cns}
\end{equation}
This implies that
\begin{eqnarray}
  g_{k+1} (\bar{a}_k, x; \beta)
  & = &
  P(Y_{k+1}^{\bar{a}_k} = 0 \mid X=x, Y_k^{\bar{a}_k} = 1)
  \nonumber \\
  & = &
  F_{T^{\bar{a}_k}} (k+1 \mid X=x, T^{\bar{a}_k} \geq k)
  \nonumber \\
  & = &
1 - \exp \left \{ - \int_k^{k+1} \lambda^{\bar{a}_K} (s \mid X=x; \; \beta) \; ds \right\}.
\label{eq:g.defineT}
\end{eqnarray}
For simplicity, we shall assume $\lambda^{\bar{0}_K} (t \mid X=0) = \lambda_{k0}$ is constant over $t \in (k, k+1]$.

In Sections~\ref{sect:K1} and~\ref{sect:manyAX}, we assumed $P(Y_{k+1}^{\bar{a}_k} = 0 \mid X=x, B=b, \bar{L}_k^{\bar{a}_{k-1}} = \bar{l}_k, Y_k^{\bar{a}_{k-1}} = 1)$ depends on $x$, $b$ and $\bar{l}_k$ only through $x$ and the risk score function $h_k^{\bar{a}_k} (x, b, \bar{l}_k)$.
Now we assume this is also true of $\lambda^{\bar{a}_k} (t \mid x, b, \bar{l}_k)$, the potential hazard at time $t$ $(k < t \leq k+1)$ given $X=x$, $B=b$ and $\bar{L}_k^{\bar{a}_{k-1}} = \bar{l}_k$ when we intervene to set $\bar{A}_k = \bar{a}_k$.
It then follows from equations~(\ref{eq:FT.cns}) and~(\ref{eq:g.defineT}) that, when $0 \leq u \leq g_{k+1} (\bar{a}_k, x; \beta)$,
\begin{eqnarray*}
  F_{T^{\bar{a}_k}}^{-1} (k + u \mid X=x, Y_k^{\bar{a}_k} = 1)
  & = &
        \frac{ \log(1-u) }{ \log \{ 1 - g_{k+1} (\bar{a}_k, x; \beta) \} }.
\end{eqnarray*}

This means we can use the algorithm in Section~\ref{sect:generalK} with $g_k (\bar{a}_{k-1}, X; \beta)$ defined by equation~(\ref{eq:g.defineT}) and the following additional step:

9.  If $Y_{k+1} = 0$, calculate $T = k + \log(1-U_{Y_{k+1}^{\bar{a}_k}}) / \log \{ 1 - g_{k+1}(\bar{a}_k, X; \beta) \}$.

This step means that when the sampled individual fails between times $k$ and $k+1$, its continuous failure time is calculated from $U_{Y_{k+1}^{\bar{a}_k}}$.
If one were instead using the extended algorithm of Section~\ref{sect:manyAX}, its step 10 would be modified to:

10.  If $Y_{k+1,1} = 0$, calculate $T_1 = k + \log(1-U_{Y_{k+1,1}^{\bar{a}_k}}) / \log \{ 1 - g_{k+1}(\bar{a}_k, X_1; \beta) \}$ and then stop.
If $k=K$, stop.

Note that for the Cox MSM, equation~(\ref{eq:g.defineT}) becomes
\[
g_{k+1} (\bar{a}_k, x; \beta) =
  1 - \exp \left[
        -  \exp \left\{
        \log \lambda_{k0}
     + \sum_{j=1}^J \beta_{kj} q_{kj} (\bar{a}_k, x)
     \right\} \right],
\]
which corresponds to a MSM for a discrete failure time with complementary log log link function.

\section{Illustrative simulation study}

\label{sect:simulationstudy}

We now illustrate how the proposed methods could be used in practice to carry out a simulation study.
In this illustrative study, the aim is to compare the simple sandwich variance estimator and non-parametric bootstrap as two methods for calculating CIs and testing for treatment effect modification when fitting an MSM using IPTW.
For this, we wish to generate data under a specified Cox MSM, which is then fitted to these data via IPTW using a correctly-specified model for the weights.
Unlike the simple sandwich estimator, bootstrap accounts for the uncertainty associated with estimating the IPT weights.

\subsection{Methods}

Data for 10 visits ($K=9)$, binary treatment $A_k$ and continuous failure time $T$ were simulated as follows.

We chose to have two independent baseline covariates $X = (X_1, X_2)^\top$ in the MSM, and two baseline variables $B = (B_1, B_2)^\top$ that are not in the MSM, and assume $B_1$ and $B_2$ are conditionally independent given $X$.
Further, $L_k = (L_{k1}, L_{k2})^\top$ are two time-dependent variables that are conditionally independent given $(X, B, \bar{L}_{k-1}, \bar{A}_{k-1})$ and $Y_k=1$ ($k=0, \ldots, 9$).
We specify: $X_1 \sim \mbox{Normal} (0,1)$; $X_2 \sim \mbox{Bernoulli} (0.5)$; $B_1 \mid X \sim \mbox{Normal} (-0.2 + 0.4 X_2, 1)$; $B_2 \mid X \sim \mbox{Normal} (0.2 X_1, 1)$; $L_{01} \mid X, B \sim \mbox{Normal} (0.2 X_1, 1)$; $L_{02} \mid X, B \sim \mbox{Bernoulli} ( \mbox{expit} (-0.2 + 0.4 X_2) )$; $L_{k1} \mid X, B, \bar{L}_{k-1}$, $\bar{A}_{k-1}, Y_k=1 \sim \mbox{Normal} (0.3 + 0.4 B_2 + 0.7 L_{k-1,1} - 0.6 A_{k-1}, 1)$ and $L_{k2} \mid X, B, \bar{L}_{k-1}, \bar{A}_{k-1}$, $Y_k=1 \sim \mbox{Bernoulli} ( \mbox{expit} (-0.2 + 0.4 B_2 + L_{k-1,2} - 0.6 A_{k-1}) )$ for $k=1, \ldots, 9$; and 
$
P( A_k = 1 \mid X, B, \bar{L}_k, \bar{A}_{k-1}, Y_k=1 )
=
\mbox{expit} ( -1 +
$
$
\delta_1 X_1 + \delta_2 X_2 + \delta_3 B_1 + \delta_4 L_{k1} + \delta_5 L_{k2} + A_{k-1} )
$ for $k=0, \ldots, 9$, where we consider two values for $\delta = (\delta_1,\delta_2,\delta_3,\delta_4,\delta_5)$: $\delta_{\rm low} = (0.1,0.15,0.1,0.3,0.3)$ and $\delta_{\rm high} = (0.2,0.3,0.2,0.6,0.6)$.

The algorithm of Section~\ref{sect:manyAX} with the modified step 10 of Section~\ref{sect:cns.time} was used to generate data satisfying the following Cox MSM:
$\lambda^{\bar a_k} (t \mid X) = \lambda_{k0} \exp ( \beta_1 X_1 + \beta_2 X_2$ $+ \beta_3 a_k + \beta_4 X_1 a_k )$
with $\beta = (\beta_1,\beta_2,\beta_3,\beta_4)=(0.5,0.5,-1,-0.4)$ and $\lambda_{k0}=\exp(-3.3)$.
Note that the interaction term $\beta_4 X_1 a_k$ with $\beta_4 \neq 0$ makes $X_1$ a treatment effect modifier.

We chose risk score function $h_k^{\bar{a}_k} (x, b, \bar{l}_k) = 0.3 b_1 + 0.5 b_2 + l_{k1} + l_{k2}$.
This means that when $\bar{A}_K$ is set to $\bar{a}_K$ by an intervention, an individual with large values of $B_1$, $B_2$, $L_{k1}$ and $L_{k2}$ is more likely to fail at time $k+1$ than is the average individual with the same value of $X$.
We consider two values for the copula correlation $\rho_k$: $\rho_{\rm low} = -0.5$ (weak dependence) and $\rho_{\rm high} = -0.9$ (strong dependence).

We see that $X_1$, $X_2$, $B_1$, $L_{k1}$ and $L_{k2}$ influence both $A_k$ and $Y_{k+1}^{\bar{a}_k}$, and are hence confounders.
Confounding is strongest when $\rho = \rho_{\rm high}$ and $\delta = \delta_{\rm high}$, and is weakest when $\rho = \rho_{\rm low}$ and $\delta = \delta_{\rm low}$.
Variable $B_2$ is a common cause of $L_k$ and $Y_{k+1}^{\bar{a}_k}$ but not a confounder, because it does not (directly) influence $A_k$.
We chose to make $B_2$ an unobserved variable; our algorithm generates values for $B_2$ but these are not used in the data analysis.

We generated random censoring times from an exponential distribution with rate $\exp(-3.6)$.
The marginal probability of observing an individual's failure time varied from 0.21 to 0.25, depending on the values of $\rho$ and $\delta$.
The marginal probabilities of (random) censoring before time 10 and of administrative censoring at time 10 varied, respectively, from 0.20 to 0.21 and from 0.54 to 0.59.

We considered 12 scenarios, corresponding to the two values of each of $\rho$ and $\delta$ and three sample sizes: $n=250$, 500 and 1000.
For each scenario we generated 1000 data sets.
The Cox MSM
$
\lambda^{\bar a_k}(t \mid X) = \lambda_{0}(t)\exp\left(\beta_1 X_1 + \beta_2 X_2 + \beta_3 a_k + \beta_4 X_1 a_k + \beta_5 X_2 a_k\right)
$
was fitted to each data set using either no weights (`naive method') or IPTW.
This MSM is correctly specified but includes the redundant interaction $X_2 a_k$, whose coefficient $\beta_5$ equals zero.
When using IPTW, (stabilised) weights at visit $k$ were calculated in the standard way\cite{Hernan2000} as
\begin{equation}
\prod_{j=0}^k \frac{ \hat{p} (A_j \mid X, \bar{A}_{j-1}, Y_j=1) }
{ \hat{p} (A_j \mid X, B_1, \bar{L}_j, \bar{A}_{j-1}, Y_j=1) },
\label{eq:stable.weights}
\end{equation}
with the terms in the denominator of expression~(\ref{eq:stable.weights}) calculated by fitting a correctly specified logistic regression model to the data pooled across all visits, allowing a separate intercept for each visit but assuming other coefficients to be the same at all visits.
The terms in the numerator of~(\ref{eq:stable.weights}) were calculated by fitting the corresponding (misspecified) model that omits $B_1$ and $\bar{L}_j$.
We calculated 95\% CIs for the $\beta$ parameters using the sandwich variance estimator (`sandwich CIs') and the percentile bootstrap method (with 1000 bootstrap samples) (`bootstrap CIs').
Size-0.05 tests of the null hypotheses of no treatment effect modification by each of $X_1$ and $X_2$ were performed by rejecting if the 95\% CI for, respectively, $\beta_4$ and $\beta_5$ excludes zero.

\subsection{Results}

Table~\ref{tab:sim.bias} shows the bias of the naive and IPTW parameter estimators.
The naive estimators are biased, particularly that of $\beta_3$.
When $n=1000$ and $\delta = \delta_{\rm low}$, IPTW estimators are approximately unbiased.
When $n=1000$ and $\delta = \delta_{\rm high}$, there is some bias in the IPTW estimator of $\beta_3$, particularly when $\rho = \rho_{\rm high}$, but this is small compared to the true value (i.e.\ $\beta_3 = -1$).
Bias in the IPTW estimator of $\beta_3$ tends to worsen slightly when $n=500$ or $n=250$, but does not exceed 0.12.

Table~\ref{tab:sim.cover} shows coverage of CIs.
Coverage of naive CIs is very low for $\beta_3$, reflecting the bias in the naive point estimate of $\beta_3$.
When $n=1000$ and $\delta = \delta_{\rm low}$, there is no obvious difference between coverage of sandwich and bootstrap CIs, or power to detect effect modification.
When $n=1000$ and $\delta = \delta_{\rm high}$, sandwich and bootstrap CIs both tend to slight undercoverage, but coverage of bootstrap CIs is closer to 95\%.
Results for $n=500$ and $n=250$ are qualitatively similar, with coverage deteriorating for both methods but remaining closer to 95\% for bootstrap CIs than for sandwich CIs.
Table~\ref{tab:sim.power} shows the power of tests to detect effect modification by $X_1$ (note that the power for $X_2$ is just one minus coverage of the CI for $\beta_5$).
Power is sometimes greater for the sandwich method than for bootstrap, but this usually reflects undercoverage of sandwich CIs.

\begin{table}
\begin{tabular}{rrrrrrr}
  \hline
  & \multicolumn{2}{c}{$n=1000$}  & \multicolumn{2}{c}{$n=500$}  & \multicolumn{2}{c}{$n=250$} \\
  \cmidrule(lr){2-3}\cmidrule(lr){4-5}\cmidrule(lr){6-7}
  &Naive&IPTW&Naive&IPTW&Naive&IPTW \\ \hline
\multicolumn{4}{l}{$\delta_{\rm low},\rho_{\rm low}$} \\
$\beta_1$ &  0.013 &  0.003 &  0.019 &  0.005 &  0.022 &  0.009 \\
$\beta_2$ &  0.011 &  0.000 &  0.019 &  0.018 &  0.016 &  0.018 \\
$\beta_3$ &  0.594 & -0.020 &  0.577 & -0.045 &  0.552 & -0.069 \\
$\beta_4$ & -0.034 & -0.005 & -0.039 &  0.001 & -0.039 & -0.005 \\
$\beta_5$ & -0.062 &  0.003 & -0.026 &  0.043 & -0.032 &  0.020 \\
  \hline
  \multicolumn{4}{l}{$\delta_{\rm high},\rho_{\rm low}$} \\
$\beta_1$ &  0.011 &  0.009 &  0.017 &  0.018 &  0.020 &  0.023 \\
$\beta_2$ &  0.003 &  0.004 & -0.002 &  0.013 &  0.004 &  0.010 \\
$\beta_3$ &  1.002 &  0.027 &  0.994 &  0.020 &  0.973 & -0.009 \\
$\beta_4$ & -0.061 & -0.010 & -0.074 & -0.012 & -0.079 & -0.025 \\
$\beta_5$ & -0.088 &  0.007 & -0.083 &  0.001 & -0.042 &  0.079 \\
  \hline
  \multicolumn{4}{l}{$\delta_{\rm low},\rho_{\rm high}$} \\
$\beta_1$ &  0.038 &  0.000 &  0.045 &  0.004 &  0.038 & -0.001 \\
$\beta_2$ &  0.038 &  0.007 &  0.047 &  0.015 &  0.041 &  0.000 \\
$\beta_3$ &  1.001 & -0.008 &  1.005 &  0.002 &  0.954 & -0.061 \\
$\beta_4$ & -0.070 & -0.003 & -0.080 & -0.014 & -0.066 &  0.000 \\
  $\beta_5$ & -0.105 & -0.006 & -0.108 & -0.007 & -0.093 &  0.035 \\
  \hline
\multicolumn{4}{l}{$\delta_{\rm high},\rho_{\rm high}$} \\
$\beta_1$ &  0.046 &  0.000 &  0.060 &  0.019 &  0.048 &  0.016 \\
$\beta_2$ &  0.026 &  0.025 &  0.039 &  0.036 &  0.052 &  0.051 \\
$\beta_3$ &  1.773 &  0.051 &  1.788 &  0.083 &  1.792 &  0.123 \\
$\beta_4$ & -0.118 & -0.003 & -0.131 & -0.013 & -0.116 & -0.018 \\
$\beta_5$ & -0.140 & -0.008 & -0.143 & -0.001 & -0.145 & -0.020 \\
  \hline
\end{tabular}
\caption{Bias in estimators of $\beta_1, \ldots, \beta_5$ from the naive and IPTW analyses for 12 scenarios: two values of $\delta$ and $\rho$ and three sample sizes $n$.  The maximum Monte Carlo SEs for the bias in the 20 naive estimators when, respectively, $n=1000$, 500 and 250 are 0.009, 0.013 and 0.018.  The corresponding values for the IPTW estimators are 0.015, 0.020 and 0.027.}
\label{tab:sim.bias}
\end{table}

\begin{table}
\begin{tabular}{lrrrrrrrrr}
  \hline
  & \multicolumn{3}{c}{$n=1000$} & \multicolumn{3}{c}{$n=500$} & \multicolumn{3}{c}{$n=250$} \\
  \cmidrule(lr){2-4}\cmidrule(lr){5-7}\cmidrule(lr){8-10}
  & Naive & Sand & Boot & Naive & Sand & Boot & Naive & Sand & Boot \\
  \hline
  \multicolumn{4}{l}{$\delta_{\rm low},\rho_{\rm low}$}\\
  $\beta_1$ & 0.943 & 0.944 & 0.945 & 0.933 & 0.926 & 0.923 & 0.942 & 0.930 & 0.939 \\ 
  $\beta_2$ & 0.947 & 0.942 & 0.945 & 0.949 & 0.935 & 0.933 & 0.971 & 0.948 & 0.947 \\ 
  $\beta_3$ & 0.216 & 0.945 & 0.938 & 0.486 & 0.946 & 0.936 & 0.685 & 0.953 & 0.940 \\ 
  $\beta_4$ & 0.937 & 0.935 & 0.938 & 0.927 & 0.931 & 0.937 & 0.936 & 0.928 & 0.938 \\ 
  $\beta_5$ & 0.941 & 0.940 & 0.943 & 0.951 & 0.939 & 0.938 & 0.967 & 0.945 & 0.948 \\ 
  \hline
  \multicolumn{4}{l}{$\delta_{\rm high},\rho_{\rm low}$}\\
  $\beta_1$ & 0.945 & 0.905 & 0.927 & 0.942 & 0.908 & 0.931 & 0.941 & 0.894 & 0.934 \\ 
  $\beta_2$ & 0.954 & 0.913 & 0.934 & 0.960 & 0.896 & 0.929 & 0.954 & 0.891 & 0.927 \\ 
  $\beta_3$ & 0.005 & 0.930 & 0.942 & 0.091 & 0.915 & 0.951 & 0.338 & 0.905 & 0.932 \\ 
  $\beta_4$ & 0.918 & 0.907 & 0.938 & 0.927 & 0.910 & 0.942 & 0.943 & 0.900 & 0.944 \\ 
  $\beta_5$ & 0.930 & 0.924 & 0.943 & 0.944 & 0.905 & 0.932 & 0.950 & 0.908 & 0.936 \\ 
  \hline
  \multicolumn{4}{l}{$\delta_{\rm low},\rho_{\rm high}$}\\
  $\beta_1$ & 0.928 & 0.950 & 0.948 & 0.918 & 0.937 & 0.942 & 0.943 & 0.924 & 0.935 \\ 
  $\beta_2$ & 0.945 & 0.957 & 0.945 & 0.948 & 0.946 & 0.940 & 0.945 & 0.947 & 0.948 \\ 
  $\beta_3$ & 0.004 & 0.947 & 0.937 & 0.081 & 0.959 & 0.945 & 0.359 & 0.947 & 0.938 \\ 
  $\beta_4$ & 0.924 & 0.945 & 0.943 & 0.921 & 0.951 & 0.947 & 0.923 & 0.931 & 0.939 \\ 
  $\beta_5$ & 0.918 & 0.943 & 0.944 & 0.938 & 0.945 & 0.944 & 0.942 & 0.933 & 0.927 \\ 
  \hline
  \multicolumn{4}{l}{$\delta_{\rm high},\rho_{\rm high}$}\\
  $\beta_1$ & 0.932 & 0.910 & 0.924 & 0.932 & 0.905 & 0.927 & 0.921 & 0.886 & 0.931 \\ 
  $\beta_2$ & 0.954 & 0.918 & 0.932 & 0.949 & 0.889 & 0.919 & 0.965 & 0.915 & 0.939 \\ 
  $\beta_3$ & 0.000 & 0.920 & 0.925 & 0.000 & 0.918 & 0.926 & 0.018 & 0.925 & 0.939 \\ 
  $\beta_4$ & 0.864 & 0.926 & 0.941 & 0.894 & 0.926 & 0.951 & 0.917 & 0.912 & 0.940 \\ 
  $\beta_5$ & 0.923 & 0.920 & 0.931 & 0.928 & 0.913 & 0.925 & 0.954 & 0.923 & 0.950 \\ 
  \hline
\end{tabular}
\caption{Coverage of naive, sandwich and bootstrap 95\% CIs for $\beta_1, \ldots, \beta_5$ for 12 scenarios: two values of $\delta$ and $\rho$ and three sample sizes $n$.
  Naive CIs are constructed as unweighted (naive) point estimate plus/minus model-based SE, sandwich CIs as IPTW point estimate plus/minus square root of sandwich variance estimate, and bootstrap CIs using IPTW and percentile method.  
  When true coverage is 0.95, Monte Carlo SEs are approximately 0.007, 0.010 and 0.014 when $n=1000$, 500 and 250, respectively.}
  \label{tab:sim.cover}
\end{table}

\begin{table}
\begin{tabular}{rrrrrrrrrrr}
  \hline
  & & \multicolumn{3}{c}{$n=1000$} & \multicolumn{3}{c}{$n=500$} & \multicolumn{3}{c}{$n=250$} \\
  \cmidrule(lr){3-5}\cmidrule(lr){6-8}\cmidrule(lr){9-11}
  & &  Naive & Sand & Boot & Naive & Sand & Boot & Naive & Sand & Boot \\
  \hline
  $\delta_{\rm low}$ & $\rho_{\rm low}$   & 0.903 & 0.760 & 0.754 & 0.658 & 0.482 & 0.479 & 0.400 & 0.322 & 0.284 \\ 
  $\delta_{\rm high}$ & $\rho_{\rm low}$    & 0.946 & 0.591 & 0.584 & 0.736 & 0.436 & 0.400 & 0.454 & 0.315 & 0.245 \\ 
  $\delta_{\rm low}$ & $\rho_{\rm high}$ & 0.965 & 0.828 & 0.824 & 0.761 & 0.580 & 0.563 & 0.448 & 0.317 & 0.286 \\ 
  $\delta_{\rm high}$ & $\rho_{\rm high}$   & 0.971 & 0.620 & 0.616 & 0.793 & 0.445 & 0.433 & 0.481 & 0.305 & 0.253 \\ 
  \hline
\end{tabular}
\caption{Power of test of null hypothesis that $\beta_4=0$ for 12 scenarios: two values of $\delta$ and $\rho$ and three sample sizes $n$.  The null is rejected if naive/sandwich/bootstrap 95\% CI excludes zero.  The maximum Monte Carlo SE for the power is 0.02, 0.02 and 0.03 when $n=1000$, 500 and 250, respectively.}
  \label{tab:sim.power}
\end{table}

The improved performance of bootstrap CIs relative to sandwich CIs agrees with previous research.
Robins et al.\ (1994)\cite{Robins1994} recommended bootstrap in a related context, saying that, in their experience, the sandwich estimator can suffer from finite-sample bias.
In the context of average treatment effect estimation, Austin (2022)\cite{Austin2022} found that the sandwich estimator underestimated the true variance unless the sample size is large, and bootstrap performed better.
We also found the sandwich estimator tended to underestimate the variance more than did bootstrap, at least when $n \geq 500$ (see Appendix~\ref{appendix:variance}).
In the context of estimating a marginal hazard ratio, Austin (2016)\cite{Austin2016} recommended bootstrap, while Mao et al.\ (2018)\cite{Mao2018} found bootstrap performed better than sandwich except when both methods performed badly.
However, in these last two simulation studies, the data-generating mechanisms appear not to have been compatible with the analysis models, due to non-collapsibility of hazard ratios, which could adversely affect the sandwich estimator.
Note that there is a sandwich variance estimator that, unlike the simple sandwich estimator used here, accounts for the uncertainty in the weights.
However, accounting for this uncertainty is known to reduce the sandwich estimate, and so would not have improved coverage in our simulation study\cite{Enders2018}.

\section{Discussion}

\label{sect:discussion}

We have proposed an algorithm for simulating data from a marginal structural logistic model, Cox MSM or additive hazards MSM.
It can be used to enable fair or neutral simulation study comparisons between causal effect estimation methods based on MSMs.
It allows more general data-generating mechanisms than previously proposed algorithms.
However, unlike Young and Tchetgen's (2014)\cite{Young2014} approach, it does not allow the conditional hazard given the history of time-dependent confounders and treatment to be directly specified, a specification which may be desirable for some simulation studies.
Nor does it generate data that simultaneously satisfy multiple models, as was the goal of Young et al.\ (2010)\cite{Young2010}.
So, those previous methods remain useful (but limited in the data-generating mechanisms they allow).

Our algorithm assumes visit times are the same for all individuals and equally spaced.
It can be generalised to unequally spaced visits, by rescaling the time axis between subsequent visits, provided that visit times remain the same for all individuals.
Although we assumed in Section~\ref{sect:cns.time} that the continuous-time hazard is constant between visits, generalisation to non-constant hazards is straightforward.
In Appendix~\ref{appendix:continuous}, we adapt our algorithm to generate data for continuous-time MSMs~\cite{Saarela2016,Ryalen2019,Ryalen2020,Dong2021}, where the times at which exposure and confounders change differ between individuals.
We have also assumed the risk score is a continuous variable, but in Appendix~\ref{appendix:discrete}, we describe simple modifications of our algorithm to handle a discrete risk score, such as would arise if all confounders were discrete.

We have considered static treatment regimes, but MSMs are also used for dynamic treatment regimes\cite{Young2011,Cain2016,Emilsson2017,Garcia-Albeniz2015,Hernan2018,Zhang2014}.
For example, Cain et al.\ (2016)\cite{Cain2016} compare two dynamic regimes for HIV-positive patients: starting treatment when patient's HIV RNA exceeds 400 copies/mL, and starting when RNA exceeds 1000 copies/mL.
Analyses of this type essentially involve assuming a MSM of the form
$
P(Y_{k+1}^s=1 \mid X, Y_k^s=1) = \mbox{expit} ( \beta_{k0} + \beta_1 X + \beta_{k2} s )
$,
where $X$ are baseline covariates and $Y_k^s$ is an indicator of survival at visit $k$ when dynamic treatment regime $S=s$ is used ($s=0,1$).
For example, $S=1$ and $S=0$ could mean start treatment when RNA exceeds, respectively, 400 and 1000 copies/mL.
The simulation approach Evans and Didelez (2024)\cite{Evans2024} describe, and which we have adopted here, may not be natural for such regimes, for the following reason.
Suppose we omit $X$ from the above MSM and choose $\beta_{00} = 5$, $\beta_{02} = 0$, $\beta_{10} = -5$ and $\beta_{12} = 10$.
Then regime $S=1$ almost guarantees survival to visit 2, whereas $S=0$ almost guarantees failure between visits 1 and 2.
For the two regimes to differ so much, the vast majority of individuals must cross the 400, but not the 1000, copies/mL threshold between visits 1 and 2.
However, our approach ostensibly allows the $\beta$ parameters to be chosen entirely separately from the parameters of the treatment model.
For example, we might choose to distribute the treatment start times of patients when assigned to regime $S=1$ uniformly over the visit times.
More research is needed to simulate data for dynamic-regime MSMs.


\subsubsection*{Acknowledgements}

We thank Jon Michael Gran of the University of Oslo for his valuable comments on an early draft of this work.
SRS is funded by UKRI (Unit programme numbers MC UU 00002/10) and supported by the National Institute for Health Research (NIHR) Cambridge Biomedical Research Centre (BRC-1215-20014).
RHK is funded by a UK Research \& Innovation Future Leaders Fellowship (MR/S017968/1).
The views expressed are those of the authors and not necessarily those of PHE, the NHS, the NIHR or the Department of Health and Social Care.

\subsubsection*{Conflict of interest}

The authors have declared no conflict of interest.

\subsubsection*{Data availability}

The data that support the findings of this study are available in the supplementary material of this article.

\subsubsection*{ORCID}

Shaun R. Seaman https://orcid.org/0000-0003-3726-5937 \\
Ruth H. Keogh https://orcid.org/0000-0001-6504-3253

\newpage

\appendix

\begin{center}
{\LARGE \bf   Supplementary Materials for `Simulating data from marginal structural models for a survival time outcome'}
\end{center}

\begin{center}
{\large \bf   Shaun R.\ Seaman and Ruth H.\ Keogh}
\end{center}

\section{Evans and Didelez's MSM simulation study}

\label{appendix:YTT}

In this appendix, we discuss the model used in Evans and Didelez's Example 6.2 and the simulation study using this model in their Appendix F.
We explain how Evans and Didelez are able to simulate data from this model and discuss the restrictiveness of the model.
We then describe how our unextended algorithm in Section 3 could very easily be modified to simulate from this same model (by replacing the copula by an odds ratio), and explain why our extended algorithm in Section 4 is not required.

Evans and Didelez's Example 6.2 comes from Young and Tchetgen Tchetgen (2014)\cite{Young2014} (henceforth, `YTT'), who consider the scenario where the failure time is discrete, there are no baseline covariates $X$ or $B$, the treatment $A_k$ is binary, there only one confounder $L_k$, which is binary, and
\begin{equation}
  P( Y_{k+1} = 0 \mid \bar{L}_k, \bar{A}_k, Y_k=1 )
  =
  P( Y_{k+1} = 0 \mid L_k, A_k, A_{k-1}, Y_k=1 )
  = s(L_k, A_k, A_{k-1})
  \label{eq:YonY} \\
\end{equation}
and
\begin{equation}
  P( L_k = 1 \mid \bar{L}_{k-1}, \bar{A}_{k-1}, Y_k=1 )
  =
  P( L_k = 1 \mid A_{k-1}, Y_k=1 )
  = r(A_{k-1})
  \label{eq:LonA}
\end{equation}
for all $k=0, \ldots, K$ and for some functions $s$ and $r$.
Equation~(\ref{eq:YonY}) means that the (discrete-time) hazard of failure depends only on the most recent treatment and confounder values.
Equation~(\ref{eq:LonA}) means that the conditional distribution of the time-dependent confounder depends only on the most recent treatment.

YTT show that in this scenario,
\begin{equation}
  \frac{ P(Y_{k+1}^{\bar{a}_k} = 0 \mid Y_k^{\bar{a}_{k-1}} = 1) }
       { P(Y_{k+1}^{\bar{0}_k} = 0 \mid Y_k^{\bar{0}_{k-1}} = 1) }
       =
       \exp( \psi_0 a_k + \psi_1 a_{k-1} + \psi_2 a_k a_{k-1} )
       \label{eq:causal}
\end{equation}
with $\psi_0$, $\psi_1$ and $\psi_2$ given by YTT's equations~(10)--(12).

More specifically, if
\begin{eqnarray}
  P( Y_{k+1} = 0 \mid \bar{L}_k, \bar{A}_k, Y_k=1 )
  & = &
  \frac{ \exp ( \theta_0 + \theta_1 L_k + \theta_2 A_k + \theta_3 A_{k-1} ) } { 1 + \exp ( \theta_0 + \theta_1 L_k + \theta_2 A_k + \theta_3 A_{k-1} ) },
  \label{eq:YonY.specific}
  \\
  P( L_k = 1 \mid \bar{L}_{k-1}, \bar{A}_{k-1}, Y_k=1 )
  & = &
  \frac{ \exp (\beta_1 A_{k-1}) } { 1 + \exp (\beta_1 A_{k-1}) }
  \label{eq:LonA.specific}
\end{eqnarray}
and failure is rare, so that equation~(\ref{eq:YonY.specific}) becomes
\begin{equation}
P( Y_{k+1} = 0 \mid \bar{L}_k, \bar{A}_k, Y_k=1 )
\approx 
\exp ( \theta_0 + \theta_1 L_k + \theta_2 A_k + \theta_3 A_{k-1} ),
\end{equation}
then $\psi_0$, $\psi_1$ and $\psi_2$ are given by YTT's equations~(22)--(24).
YTT note that this result easily generalises to the situation where the right-hand side of equation~(\ref{eq:LonA.specific}) is replaced by $\exp (\beta_0 + \beta_1 A_{k-1}) / \{ 1 + \exp (\beta_0 + \beta_1 A_{k-1}) \}$.

For the remainder of this Appendix~\ref{appendix:YTT} we shall be assuming, unless indicated otherwise, that all the assumptions stated so far in this Appendix~\ref{appendix:YTT} are satisfied.

By the consistency assumption
and equation~(\ref{eq:YonY.specific}), we have
\begin{eqnarray}
  &&
  P( Y_{k+1}^{\bar{a}_k} = 0 \mid \bar{L}_k^{\bar{a}_{k-1}} = \bar{l}_k, \bar{A}_k = \bar{a}_k, Y_k^{\bar{a}_{k-1}} = 1 )
  \nonumber \\
  && \hspace{1cm} =
  P( Y_{k+1} = 0 \mid \bar{L}_k = \bar{l}_k, \bar{A}_k = \bar{a}_k, Y_k = 1 )
  \nonumber \\
  && \hspace{1cm} =
  \frac{ \exp ( \theta_0 + \theta_1 l_k + \theta_2 a_k + \theta_3 a_{k-1} ) } { 1 + \exp ( \theta_0 + \theta_1 l_k + \theta_2 a_k + \theta_3 a_{k-1} ) }.
\label{eq:Ya.equals.Y}
\end{eqnarray}
The causal DAG (Figure~1 in our article, and see also Section 3.2 of our article) implies that
\begin{equation}
    P( Y_{k+1}^{\bar{a}_k} = 0 \mid \bar{L}_k^{\bar{a}_{k-1}}, \bar{A}_k = \bar{a}_k, Y_k^{\bar{a}_{k-1}} = 1 )
  =
  P( Y_{k+1}^{\bar{a}_k} = 0 \mid \bar{L}_k^{\bar{a}_{k-1}}, Y_k^{\bar{a}_{k-1}} = 1 ).
  \label{eq:YonL}
\end{equation}
From equations~(\ref{eq:Ya.equals.Y}) and (\ref{eq:YonL}), we have
\begin{equation}
P( Y_{k+1}^{\bar{a}_k} = 0 \mid \bar{L}_k^{\bar{a}_{k-1}}, Y_k^{\bar{a}_{k-1}} = 1 )
=
  \frac{ \exp ( \theta_0 + \theta_1 L_k^{\bar{a}_{k-1}} + \theta_2 a_k + \theta_3 a_{k-1} ) } { 1 + \exp ( \theta_0 + \theta_1 L_k^{\bar{a}_{k-1}} + \theta_2 a_k + \theta_3 a_{k-1} ) }.
\label{eq:potYonpotY}
\end{equation}
We see from equation~(\ref{eq:potYonpotY}) that the risk ranking of individuals with $Y_k^{\bar{a}_{k-1}}=1$ and different values of $\bar{L}_k^{\bar{a}_{k-1}}$ depends only on $L_k^{\bar{a}_{k-1}}$ and whether $\theta_1$ is positive or negative.
Hence, by the definition of the risk score function\footnote{Recall that there is no baseline variables $X$ or $B$ in this example.}, $h_k^{\bar{a}_k} (\bar{l}_k)$ is just $l_k$ (if $\theta_1 > 0)$ or $- l_k$ (if $\theta_1 < 0)$ or some monotonically increasing function of $l_k$ (or $- l_k$).
Suppose (without loss of generality) that $h_k^{\bar{a}_k} (\bar{l}_k) = l_k$, and so $H_k^{\bar{a}_k} = L_k^{\bar{a}_{k-1}}$.

We also see from equation~(\ref{eq:potYonpotY}) that the odds ratio between $1 - Y_{k+1}^{\bar{a}_k}$ and $L_k^{\bar{a}_{k-1}}$ given $Y_k^{\bar{a}_{k-1}} = 1$ is $\exp( \theta_1 )$.
Hence, the odds ratio between $Y_{k+1}^{\bar{a}_k}$ and $L_k^{\bar{a}_{k-1}}$ given $Y_k^{\bar{a}_{k-1}} = 1$ is $\exp( - \theta_1 )$.

By equation~(\ref{eq:LonA.specific}), the consistency assumption and the causal DAG, we have the CDF of $L_k^{\bar{a}_k}$ given $Y_k^{\bar{a}_k} = 1$:
\begin{eqnarray}
  F_{ L_k^{\bar{a}_k} } (0 \mid Y_k^{\bar{a}_k} = 1)
  & = &
  \frac{1} { 1 + \exp ( \beta_1 a_{k-1} ) }
  \nonumber \\
  F_{ L_k^{\bar{a}_k} } (1 \mid Y_k^{\bar{a}_k} = 1)
  & = &
  1.
  \label{eq:marginalL}
\end{eqnarray}

Also, if we know $\psi_* = \log P(Y_{k+1}^{\bar{0}_k} = 0 \mid Y_k^{\bar{0}_{k-1}} = 1)$, then it follows from equation~(\ref{eq:causal}) that we know the distribution of $Y_{k+1}^{\bar{a}_k}$ given $Y_k^{\bar{a}_{k-1}} = 1$:
\begin{equation}
  P(Y_{k+1}^{\bar{a}_k} = 0 \mid Y_k^{\bar{a}_{k-1}} = 1)
  =
  \exp( \psi_* + \psi_0 a_k + \psi_1 a_{k-1} + \psi_2 a_k a_{k-1} ).
\end{equation}

To recap, if we know $(\psi_*, \psi_0, \psi_1, \psi_2, \beta_1, \theta_1)$, then we know: i) the (marginal) distribution of the binary variable $Y_{k+1}^{\bar{a}_k}$ given $Y_k^{\bar{a}_{k-1}} = 1$; ii) the (marginal) distribution of the binary variable $L_k^{\bar{a}_{k-1}}$ given $Y_k^{\bar{a}_{k-1}} = 1$; and iii) odds ratio between $Y_{k+1}^{\bar{a}_k}$ and $L_k^{\bar{a}_{k-1}}$ given $Y_k^{\bar{a}_{k-1}} = 1$.

The joint distribution of $(Y_{k+1}^{\bar{a}_k}, L_k^{\bar{a}_{k-1}})$ given $Y_k^{\bar{a}_{k-1}} = 1$ follows immediately from i)--iii).
The conditional distribution of $Y_{k+1}^{\bar{a}_k}$ given $L_k^{\bar{a}_{k-1}}$ and $Y_k^{\bar{a}_{k-1}} = 1$ then follows immediately from this joint distribution, and  it is easy to simulate $(Y_{k+1}^{\bar{a}_k}, L_k^{\bar{a}_{k-1}})$ from this conditional distribution.
This is almost what Evans and Didelez do.
However, it is not quite what they do, because they actually specify $(\psi_0, \psi_1, \psi_2, \beta_1, \theta_0, \theta_1)$, rather than $(\psi_*, \psi_0, \psi_1, \psi_2, \beta_1, \theta_1)$.
However, it is straightforward to calculate $\psi_*$ from $\theta_0$ and $\theta_1$, because
\begin{eqnarray}
  &&    
  \exp( \psi_* )
  =
  P(Y_{k+1}^{\bar{0}_k} = 0 \mid Y_k^{\bar{0}_{k-1}} = 1)
 \nonumber \\
  && =
    \sum_{l_k=0}^1 P( Y_{k+1}^{\bar{0}_k} = 0 \mid L_k^{\bar{0}_{k-1}} = l_k, Y_k^{\bar{0}_{k-1}} = 1 ) \times P( L_k^{\bar{0}_{k-1}} = l_k \mid Y_k^{\bar{0}_{k-1}} = 1 )
 \nonumber \\
  && = 
  \frac{ \exp (\theta_0) } { 1 + \exp (\theta_0) }
  \times
  \frac{ 1 } { 1 + \exp (\beta_1 \times 0) }
  +
  \frac{ \exp (\theta_0 + \theta_1) } { 1 + \exp (\theta_0 + \theta_1) }
  \times
     \frac{ \exp (\beta_1 \times 0) } { 1 + \exp (\beta_1 \times 0) }
     \nonumber \\
  &&
  \label{eq:using.two.equations} \\
  && = 
  \frac{ \exp (\theta_0) } { 2 \{ 1 + \exp (\theta_0) \} }
  +
     \frac{ \exp (\theta_0 + \theta_1) } { 2 \{ 1 + \exp (\theta_0 + \theta_1) \} }.
     \nonumber
\end{eqnarray}
Note that line~(\ref{eq:using.two.equations}) follows from equations~(\ref{eq:potYonpotY}) and~(\ref{eq:marginalL}).

Although this approach of Evans and Didelez works when there are no baseline covariates $(X, B)$, $L_k$ is a single binary variable, failure is rare and equations~(\ref{eq:causal})--(\ref{eq:LonA.specific}) hold, Evans and Didelez do not say what they would do if $L_k$ were continuous and/or a vector, or when equations~(\ref{eq:causal})--(\ref{eq:LonA.specific}) do not hold.
For example, YTT also consider the corresponding scenario where, instead of being a single binary variable, $L_k$ is a single continuous variable with
$
L_k  \mid \bar{L}_{k-1}, \bar{A}_{k-1}, Y_k=1 \sim \mbox{Normal} (\beta_1 A_{k-1}, \sigma^2).
$

Our approach of using a copula to describe the association between $Y_{k+1}^{\bar{a}_k}$ and $L_k^{\bar{a}_{k-1}}$ given $Y_k^{\bar{a}_{k-1}} = 1$ would not work in the scenario of a single binary $L_k$,
because we assume that $H_k^{\bar{a}_k}$ is a continuous random variable, whereas here $H_k^{\bar{a}_k} = L_k$ is binary.
However, if we take our method and replace the copula by an odds ratio parameter, then it does exactly what Evans and Didelez have done.
This is what we label as `Option 2' in our Appendix~\ref{appendix:discrete}.
Alternatively, we could continue to use a copula but with the modification described as `Option 1' in Appendix~\ref{appendix:discrete}.

Finally, we remark that in the scenario that Evans and Didelez consider, $H_k^{\bar{a}_k} = L_k^{\bar{a}_k}$ and $F_{ H_k^{\bar{a}_k} } (h \mid Y_k^{\bar{a}_k} = 1)$ is a known function (see equation~(\ref{eq:marginalL})).
This remains true in the more general scenario where equation~(\ref{eq:YonY}) holds and $L_k$ can be discrete or continuous and can be scalar or vector but obeys
\[
  p ( L_k \mid \bar{L}_{k-1}, \bar{A}_{k-1}, Y_k=1 )
  =
  p ( L_k \mid \bar{A}_{k-1}, Y_k=1 ).
\]
Recall that when $F_{ H_k^{\bar{a}_k} } (h \mid Y_k^{\bar{a}_k} = 1)$ is a known function, there is no need to estimate it using the `extended algorithm' described in Section 4 of our article.

\section{Proof of equation in Section~\ref{sect:sample.Y2}}

\label{appendix:proof}

By the consistency assumption,
\begin{eqnarray}
  &&
  P(Y_2^{\bar{a}_1} = 0 \mid X, B, L_0, \bar{A}_1 = \bar{a}_1, Y_1=1, L_1)
  \nonumber \\
  && \hspace{1cm} =
        P(Y_2^{\bar{a}_1} = 0 \mid X, B, L_0, A_0=a_0, A_1^{a_0} = a_1, Y_1^{a_0} =1, L_1^{a_0}).
     \label{eq:consist.exch2}
\end{eqnarray}
%

\begin{figure}
\begin{center}
\begin{tikzpicture}[auto, node distance=2cm, thick, node/.style={font=\sffamily\Large}]
  \node[node] (BX) {$(B, X)$};
  \node[node] (A0) [right = 3cm of BX] {$A_0$};
  \node[node] (A0mida0) [right = -0.25cm of A0] {$\mid$};
  \node[node] (a0) [right = -0.25cm of A0mida0] {$a_0$};
  \node[node] (A1) [right = 3cm of a0] {$A_1^{a_0}$};
  \node[node] (A1mida1) [right = -0.25cm of A1] {$\mid$};
  \node[node] (a1) [right = -0.25cm of A1mida1] {$a_1$};
  \node[node] (L0) [above left = 1.5cm and 0.5cm of A0] {$L_0$};
  \node[node] (L1) [above left = 1.5cm and 0.5cm of A1] {$L_1^{a_0}$};
  \node[node] (Y1) [below right = 1.5cm and 0.5cm of A0] {$Y_1^{a_0}$};
  \node[node] (Y2) [below right = 1.5cm and 0.5cm of A1] {$Y_2^{\bar{a}_1}$};
  \path[every node/.style={font=\sffamily\small}]
  (BX) edge[->] node [right] {} (L0)
  (BX) edge[->] node [right] {} (A0)
  (BX) edge[->,bend right=10] node [right] {} (Y1)
  (BX) edge[->,bend left=50] node [right] {} (L1)
  (BX) edge[->,bend right=35] node [right] {} (Y2)
  (BX) edge[->,bend left=30] node [right] {} (A1)
  (a0) edge[->] node [right] {} (A1)
  (a0) edge[->] node [right] {} (L1)
  (a0) edge[->] node [right] {} (Y1)
  (a0) edge[->] node [right] {} (Y2)
  (a1) edge[->] node [right] {} (Y2)
  (L0) edge[->] node [right] {} (A0)
  (L0) edge[->] node [right] {} (A1)
  (L0) edge[->] node [right] {} (L1)
  (L0) edge[->,bend right=25] node [right] {} (Y1)
  (L0) edge[->,bend left=5] node [right] {} (Y2)
  (L1) edge[->] node [right] {} (A1)
  (L1) edge[->,bend left=45] node [right] {} (Y2)
  (Y1) edge[->] node [right] {} (A1)
  (Y1) edge[->] node [right] {} (Y2)
  (Y1) edge[->] node [right] {} (L1);
\end{tikzpicture}
\end{center}
\caption{Single World Intervention Graph (SWIG) representing intervention to set $A_0 = a_0$ and $A_1 = a_1$.}
\label{fig:swig}
\end{figure}
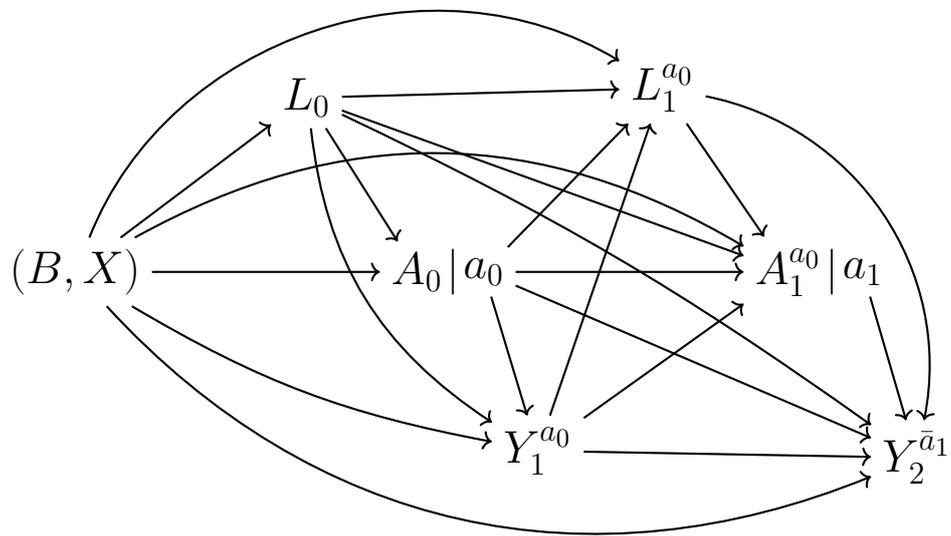

Figure~\ref{fig:swig} shows the Single World Intervention Graph (SWIG)\cite{Richardson2013} that corresponds to the causal DAG of Figure~\ref{fig:causalDAG} after intervening to set $(A_0, A_1) = (a_0, a_1)$.
By applying d-separation (see Section 3.5.2 of \cite{Richardson2013}) to this SWIG, it can be seen that
\begin{equation}
Y_2^{\bar{a}_1} \ci (A_0, A_1^{a_0}) \mid X, B, L_0, Y_1^{a_0} = 1, L_1^{a_0}.
\label{eq:dsep}
\end{equation}
It follows from equation~(\ref{eq:consist.exch2}) and expression~(\ref{eq:dsep}) that
\[
  P(Y_2^{\bar{a}_1} = 0 \mid X, B, L_0, \bar{A}_1=a_1, Y_1=1, L_1) = P(Y_2^{\bar{a}_1} = 0 \mid X, B, L_0, Y_1^{a_0} = 1, L_1^{a_0}).
\]

\section{Estimating CDF of risk score when $(X, \bar{A}_K)$ can take few values}

\label{appendix:knownF}

As mentioned in Section~\ref{sect:manyAX}, when the number of possible values of $(X, \bar{A}_K)$ is finite and fairly small, the CDF $F_{H_k^{\bar{a}_k}} (h \mid x, Y_k^{\bar{a}_{k-1}} = 1)$ of the risk score can be estimated separately for each of these possible values by a Monte Carlo procedure prior to applying the algorithm in Section~\ref{sect:generalK}.
Here we describe that Monte Carlo procedure.

Let $m$ be a large number, e.g.\ $m=10^6$.
For each possible value $(x, \bar{a}_K)$ of $(X, \bar{A}_K)$, follow these steps.
\begin{enumerate}
\item
  For $j=1, \ldots, m$, sample $B_j$ from $p(B_j \mid X_j=x)$.
  Set $k=0$.
\item
  For $j=1, \ldots, m$, sample $L_{kj}$ from $p(L_{kj} \mid X_j=x, B_j, \bar{L}_{k-1,j}, \bar{A}_{k-1,j} = \bar{a}_{k-1}, Y_{kj}=1)$.
\item
  For $j=1, \ldots, m$, calculate $H_{kj} = H_k^{\bar{a}_k} (x, B_j, \bar{L}_{kj})$.
  Let $\hat{F}_{H_k^{\bar{a}_k}} (h \mid x, Y_k^{\bar{a}_{k-1}} = 1)$ denote the empirical distribution of $\{H_{k1}, \ldots, H_{km} \}$.
\item
  If $k=K$, stop.
\item
  Let $R_{kj}$ denote the rank of $H_{kj}$ among the set $\{H_{k1}, \ldots, H_{km} \}$.
\item
  Generate $m$ independent Uniform$(0,1)$ random variables, $W_{k1}, \ldots, W_{km}$.
  For $j=1, \ldots, m$, calculate $U_{H_k j} = (R_{kj} - W_{kj}) / m$.
\item
  For $j=1, \ldots, m$, calculate $Z_{H_k^{\bar{a}_k}, j} = \Phi^{-1} (U_{H_k^{\bar{a}_k}, j})$.
\item
  For $j=1, \ldots, m$,   Sample $Z_{Y_{k+1}^{\bar{a}_k}, j} \sim \mbox{Normal} (\rho_k Z_{H_k^{\bar{a}_k}, j}, 1 - \rho_k^2)$ and calculate $U_{Y_{k+1}^{\bar{a}_k}, j} = \Phi ( Z_{Y_{k+1}^{\bar{a}_k}, j} )$.
\item
  For $j=1, \ldots, m$, set $Y_{k+1,j}=0$ if $U_{Y_{k+1},j} < g_{k+1} (\bar{a}_{k-1}, x; \beta)$ and set $Y_{k+1,j}=1$ otherwise.
\item
  Replace each individual $j$ that has $Y_{k+1,j} = 0$ with a randomly chosen individual $j^*$ ($2 \leq j^* \leq m$) that has $Y_{k+1,j^*} = 1$.
  That is, randomly generate $j^*$ and then set $B_j = B_{j^*}$, $\bar{L}_{kj} = \bar{L}_{kj^*}$ and $Y_{k+1,j} = 1$.
\item
  Let $k=k+1$ and return to step 2.
\end{enumerate}

Now $\hat{F}_{H_k^{\bar{a}_k}} (h \mid x, Y_k^{\bar{a}_{k-1}} = 1)$ (calculated at step 3) is an estimate of $F_{H_k^{\bar{a}_k}} (h \mid x, Y_k^{\bar{a}_{k-1}} = 1)$ that can be used in the algorithm of Section~\ref{sect:generalK}.

In practice, one might store only a fixed number of quantiles of this empirical distribution function and then interpolate between these when using the algorithm of Section~\ref{sect:generalK}.
For example, these could be the $(0.00001, 0.00002, \ldots 0.0001)$th, the $(0.0002, 0.0003, \ldots 0.1)$th and the $(0.101, 0.102, 0.103, \ldots, 0.5)$th quantiles (a total of $10+999+400 = 1409$ quantiles) and the corresponding 1408 quantiles between 0.501 and 0.99999.

\section{Empirical study}

\label{appendix:empirical.study}

In this section, we demonstrate that the extended algorithm does indeed generate data compatible with a pre-specified MSM.

We let $K=9$ and consider a binary treatment $A_k$.
We choose to have two independent baseline covariates $X = (X_1, X_2)^\top$ in the MSM, and two baseline variables $B = (B_1, B_2)^\top$ that are not in the MSM, and assume that $B_1$ and $B_2$ are conditionally independent given $X$.
Further, $L_k = (L_{k1}, L_{k2})^\top$ are two time-dependent variables that are conditionally independent given $(X, B, \bar{L}_{k-1}, \bar{A}_{k-1})$ and $Y_k=1$ ($k=0, \ldots, 9$).
We specify that
\begin{eqnarray*}
  X_1 & \sim & \mbox{Normal} (0,1)
  \\
  X_2 & \sim & \mbox{Bernoulli} (0.5)
  \\
  B_1 \mid X & \sim & \mbox{Normal} (-0.2 + 0.4 X_2, 1)
  \\
  B_2 \mid X & \sim & \mbox{Normal} (0.2 X_1, 1)
  \\
  L_{01} \mid X, B & \sim & \mbox{Normal} (0.2 X_1, 1)
  \\
  L_{02} \mid X, B & \sim & \mbox{Bernoulli} ( \mbox{expit} (-0.2 + 0.4 X_2) )
\end{eqnarray*}
and
\begin{eqnarray*}
  L_{k1} \mid X, B, \bar{L}_{k-1}, \bar{A}_{k-1}, Y_k=1 & \sim & \mbox{Normal} (0.3 + 0.4 B_2 + 0.7 L_{k-1,1} - 0.6 A_{k-1}, 1)
  \\
  L_{k2} \mid X, B, \bar{L}_{k-1}, \bar{A}_{k-1}, Y_k=1 & \sim & \mbox{Bernoulli} ( \mbox{expit} (-0.2 + 0.4 B_2 + L_{k-1,2} - 0.6 A_{k-1}) )
\end{eqnarray*}
for $k=1, \ldots, 9$.
We want to simulate data compatible with the following marginal structural logistic model:
\begin{equation*}
  P( Y_{k+1}^{\bar{a}_k} = 0 \mid X, Y_k^{\bar{a}_k}=1 ) = \mbox{expit} (\beta_{k0} + \beta_{k1} X_1 + \beta_{k2} X_2 + \beta_{k3} a_k)
  \hspace{1cm} (k=0, \ldots, 9).
  \label{eq:MSM.logistic}
\end{equation*}
We choose the risk score function to be $h_k^{\bar{a}_k} (x, b, \bar{l}_k) = 0.3 b_1 + 0.5 b_2 + l_{k1} + l_{k2}$ and the correlation parameter of the Gaussian copula to be $\rho_k = -0.9$.
This means that, under the interventional regime (i.e.\ when $\bar{A}_K$ is set equal to $\bar{a}_K$), an individual with large values of $B_1$, $B_2$, $L_{k1}$ and $L_{k2}$ is more likely to fail at time $k+1$ than is the average individual with the same value of $X$.

For the treatment, we specify that
\begin{eqnarray}
  &&
  P( A_k = 1 \mid X, B, \bar{L}_k, \bar{A}_{k-1}, Y_k=1 )
  =
  \nonumber \\
  && \hspace{0.5cm}
  \mbox{expit} ( -1 + 0.2 X_1 + 0.3 X_2 + 0.2 B_1 + 0.6 L_{k1} + 0.6 L_{k2} + A_{k-1} ).
   \label{eq:treatment.model}
\end{eqnarray}

It can be seen that $X_1$, $X_2$, $B_1$, $L_{k1}$ and $L_{k2}$ influence both $A_k$ and $Y_{k+1}^{\bar{a}_k}$, and are hence confounders.
The variable $B_2$ is a common cause of $L_k$ and $Y_{k+1}^{\bar{a}_k}$ but not a confounder, because it does not (directly) influence $A_k$.
We shall regard $B_2$ as an unobserved variable; the algorithm will generate values for $B_2$ but we shall discard these data.

The true values of the causal log odds ratio parameters in the MSM were chosen to be $(\beta_{k1}, \beta_{k2}, \beta_{k3}) = (0.5, 0.5, -1)$.
Three sets of values were considered for the baseline log odds parameters: $\beta_{k0} = -4.1$, $\beta_{k0} = -2.5$ and $\beta_{k0} = -1.2$.
When $\beta_{k0} = -4.1$, approximately 90\% of individuals survive to time $K+1=10$, i.e.\ $P(Y_{10}=1) \approx 0.9$.
Approximately 50\% survive when $\beta_{k0} = -2.5$, and approximately 10\% survive when $\beta_{k0} = -1.2$.

All the quantities needed to determine the data-generating mechanism have now been specified.
For each of the three values of $\beta_{k0}$, we used the algorithm in Section~\ref{sect:manyAX} to generate a sample of size $n=10^6$ individuals.

To each of the three samples, we fitted the following MSM twice, once without weighting the data and once using inverse probability of treatment (stabilised) weights:
\begin{eqnarray*}
  &&
  P( Y_{k+1}^{\bar{a}_k} = 0 \mid X, Y_k^{\bar{a}_k}=1 ) =
  \\
  && \hspace{0.5cm}
  \mbox{expit} (\beta_{k0} + \beta_1 X_1 + \beta_2 X_2 + \beta_3 a_k + \beta_4 X_1 k + \beta_5 X_2 k + \beta_6 a_k k)
  \\
  && \hspace{8cm} (k=0, \ldots, 9).
\end{eqnarray*}
Obviously, the true values of the parameters are $\beta_1 = 0.5$, $\beta_2 = 0.5$, $\beta_3 = -1$ and $\beta_4 = \beta_5 = \beta_6 = 0$.
The weights at visit $k$ were calculated in the standard way\cite{Hernan2000} as
\begin{equation}
\prod_{j=0}^k \frac{ \hat{p} (A_j \mid X, \bar{A}_{j-1}, Y_j=1) }
{ \hat{p} (A_j \mid X, B_1, \bar{L}_j, \bar{A}_{j-1}, Y_j=1) },
\label{eq:weights}
\end{equation}
with the terms in the denominator of expression~(\ref{eq:weights}) calculated by fitting the correctly specified logistic regression model for $A_j$ given $X, B_1, \bar{L}_j, \bar{A}_{j-1}$ and $Y_j=1$ (see equation~(\ref{eq:treatment.model})) to the sample.
The terms in the numerator were calculated by fitting the corresponding (misspecified) logistic regression model that omits $B_1$ and $\bar{L}_j$.

Table~\ref{tab:mslm.50} shows the point estimates obtained when $\beta_{k0} = -2.5$, i.e.\ when 50\% of individuals fail during follow-up.
Also shown are standard errors (SEs) of these estimates, calculated using either the Fisher information (for the point estimates obtained without weighting) or the sandwich variance estimator (for the point estimates obtained using IPTW).
As expected, the unweighted point estimates are biased, particularly that of $\beta_3$.
The point estimates obtained using IPTW, on the other hand, are very close to the true values of the parameters.
In particular, all estimates are within two SEs of the true parameter values.
Analogous results for 10\% failure and 90\% failure are given in Tables~\ref{tab:mslm.10} and~\ref{tab:mslm.90}.
In both cases, all the estimates are very close to the true parameter values.
These results verify that our algorithm is generating data compatible with the chosen MSM.
 \begin{table}
   \begin{tabular}{crrrrr}
&& \multicolumn{2}{c}{Unweighted} & \multicolumn{2}{c}{Weighted} \\ 
Parameter     & True & Est    & SE     & Est    & SE     \\ \hline
$\beta_{00}$  & -2.5 & -3.023 &  0.006 & -2.500 &  0.007 \\
$\beta_{10}$  & -2.5 & -3.185 &  0.006 & -2.489 &  0.007 \\
$\beta_{20}$  & -2.5 & -3.230 &  0.005 & -2.512 &  0.007 \\
$\beta_{30}$  & -2.5 & -3.220 &  0.005 & -2.490 &  0.008 \\
$\beta_{40}$  & -2.5 & -3.216 &  0.005 & -2.498 &  0.009 \\
$\beta_{50}$  & -2.5 & -3.205 &  0.006 & -2.505 &  0.010 \\
$\beta_{60}$  & -2.5 & -3.200 &  0.006 & -2.503 &  0.012 \\
$\beta_{70}$  & -2.5 & -3.182 &  0.006 & -2.479 &  0.018 \\
$\beta_{80}$  & -2.5 & -3.176 &  0.007 & -2.475 &  0.016 \\
$\beta_{90}$  & -2.5 & -3.164 &  0.007 & -2.503 &  0.017 \\
$\beta_1$    &  0.5 &  0.429 &  0.003 &  0.496 &  0.004 \\
$\beta_2$    &  0.5 &  0.403 &  0.005 &  0.497 &  0.007 \\
$\beta_3$    & -1.0 &  0.255 &  0.005 & -1.002 &  0.007 \\
$\beta_4$    &  0.0 &  0.011 &  0.001 &  0.001 &  0.001 \\
$\beta_5$    &  0.0 &  0.014 &  0.001 & -0.001 &  0.002 \\
$\beta_6$    &  0.0 &  0.042 &  0.001 &  0.003 &  0.002
   \end{tabular}
   \caption{Estimates of parameters in MSM obtained with and without inverse probability of treatment weighting, along with estimated SEs, when the marginal probability of failure before time 10 equals 0.5.}
\label{tab:mslm.50}
\end{table}
 \begin{table}
   \begin{tabular}{crrrrr}
&& \multicolumn{2}{c}{Unweighted} & \multicolumn{2}{c}{Weighted} \\
Parameter     & True & Est    & SE     & Est    & SE     \\ \hline
$\beta_{00}$  & -4.1 & -4.845 &  0.012 & -4.106 &  0.024 \\
$\beta_{10}$  & -4.1 & -5.114 &  0.013 & -4.100 &  0.022 \\
$\beta_{20}$  & -4.1 & -5.239 &  0.012 & -4.141 &  0.022 \\
$\beta_{30}$  & -4.1 & -5.273 &  0.012 & -4.101 &  0.027 \\
$\beta_{40}$  & -4.1 & -5.336 &  0.011 & -4.086 &  0.040 \\
$\beta_{50}$  & -4.1 & -5.387 &  0.012 & -4.101 &  0.042 \\
$\beta_{60}$  & -4.1 & -5.417 &  0.012 & -4.077 &  0.047 \\
$\beta_{70}$  & -4.1 & -5.481 &  0.013 & -4.017 &  0.099 \\
$\beta_{80}$  & -4.1 & -5.508 &  0.014 & -4.069 &  0.060 \\
$\beta_{90}$  & -4.1 & -5.579 &  0.016 & -3.908 &  0.126 \\
$\beta_1$     &  0.5 &  0.422 &  0.005 &  0.533 &  0.015 \\
$\beta_2$     &  0.5 &  0.364 &  0.011 &  0.523 &  0.033 \\
$\beta_3$     & -1.0 &  0.596 &  0.012 & -1.022 &  0.024 \\
$\beta_4$     &  0.0 &  0.001 &  0.001 & -0.007 &  0.005 \\
$\beta_5$     &  0.0 &  0.009 &  0.002 & -0.014 &  0.012 \\
$\beta_6$     &  0.0 &  0.081 &  0.002 &  0.006 &  0.008
   \end{tabular}
   \caption{Estimates of parameters in MSM obtained with and without inverse probability of treatment weighting, along with estimated SEs, when the marginal probability of failure before time 10 equals 0.1.}
\label{tab:mslm.10}
\end{table}
 \begin{table}
   \begin{tabular}{crrrrr}
&& \multicolumn{2}{c}{Unweighted} & \multicolumn{2}{c}{Weighted} \\ 
Parameter     & True & Est    & SE     & Est    & SE     \\ \hline
$\beta_{00}$  & -1.2 & -1.572 &  0.003 & -1.197 &  0.004 \\
$\beta_{10}$  & -1.2 & -1.661 &  0.004 & -1.205 &  0.004 \\
$\beta_{20}$  & -1.2 & -1.639 &  0.004 & -1.201 &  0.004 \\
$\beta_{30}$  & -1.2 & -1.593 &  0.004 & -1.204 &  0.005 \\
$\beta_{40}$  & -1.2 & -1.552 &  0.004 & -1.206 &  0.007 \\
$\beta_{50}$  & -1.2 & -1.507 &  0.005 & -1.199 &  0.009 \\
$\beta_{60}$  & -1.2 & -1.477 &  0.005 & -1.212 &  0.014 \\
$\beta_{70}$  & -1.2 & -1.445 &  0.006 & -1.211 &  0.017 \\
$\beta_{80}$  & -1.2 & -1.430 &  0.007 & -1.229 &  0.019 \\
$\beta_{90}$  & -1.2 & -1.394 &  0.008 & -1.184 &  0.023 \\
$\beta_1$     &  0.5 &  0.446 &  0.002 &  0.500 &  0.002 \\
$\beta_1$     &  0.5 &  0.422 &  0.004 &  0.495 &  0.005 \\
$\beta_1$     & -1.0 & -0.012 &  0.004 & -1.002 &  0.005 \\
$\beta_1$     &  0.0 &  0.013 &  0.001 &  0.001 &  0.001 \\
$\beta_1$     &  0.0 &  0.017 &  0.001 &  0.004 &  0.002 \\
$\beta_1$     &  0.0 &  0.044 &  0.001 &  0.003 &  0.002
   \end{tabular}
   \caption{Estimates of parameters in MSM obtained with and without inverse probability of treatment weighting, along with estimated SEs, when the marginal probability of failure before time 10 equals 0.9.}
\label{tab:mslm.90}
\end{table}

\section{Sensitivity of simulated failure times to the number of matches}

\label{appendix:sensitivity}

In Appendix~\ref{appendix:empirical.study}, 4999 matches were used to estimate the CDF of the risk score.
Since the data for these matches are generated randomly, a different set of 4999 matches, or a smaller number of matches, would yield a different estimate of this CDF.
This different estimate might then affect the failure times generated for the $n$ sampled individuals.
To assess how sensitive these $n$ failure times are to this estimation uncertainty, we carried out the following investigation.

Using the same data-generating mechanism as in the empirical study, we generated a value of $(B, X, \bar{L}_{10}, \bar{A}_{10})$ for each of $n=100000$ (one hundred thousand) individuals.
Just as in the extended algorithm of Section 4, this was done by sampling $L_k$ from $p(L_k \mid X, B, \bar{L}_{k-1}, \bar{A}_{k-1}, Y_k=1)$ and $A_k$ from $p(A_k \mid X, B, \bar{L}_k, \bar{A}_{k-1}, Y_k=1)$ (for $k=0, \ldots, 10$).
We also simulated values of $W_{k1} \sim \mbox{Uniform} (0,1)$ ($k=1, \ldots, 10$) for each of the 100000 individuals.
These $W_{k1}$'s were to be used in step 6 of the extended algorithm.
In addition, we simulated values of $S_{k1} \sim \mbox{Normal} (0, 1 - \rho_k^2)$ (for $k=1, \ldots, 10$) for each of the 100000 individuals.
These were to be used in step 8 of the extended algorithm (see later for how they were used).
These 100000 values of $(B, X, \bar{L}_{10}, \bar{A}_{10}, W_{k1}, \ldots, W_{k,10}, S_{k1}, \ldots, S_{k,10})$ were generated only once and then used throughout this investigation.

Next, for each of five values of $m$ (see Table~\ref{tab:m.values}), we used a slightly modified version of the extended algorithm to generate a failure time for each of the $n=100000$ sampled individuals.
\begin{table}
  \begin{tabular}{rrr}
    $m$ & min.\ $I_j$ & large $m$ \\ \hline
    10 & 0 & -- \\
    100 & 10 & 2000 \\
    500 & 50 & 10000 \\
    1000 & 100 & 20000 \\
    5000 & 500 & 100000 \\
  \end{tabular}
  \caption{The five values of $m$ that were used for this investigation, together with their corresponding thresholds for the minimum number of unique $I_j$ values allowed, and the larger value of $m$ that is used if this threshold is breached.}
  \label{tab:m.values}
\end{table}
The reason for this modification was to ensure that the only reason for changes in the generated failure times of the $n=100000$ sampled individuals was the data generated for the matches.
The modification was that:
\begin{itemize}
\item
  Step 3 was only applied to $j=2, \ldots, m$, i.e.\ the matches, because we had already generated $\bar{L}_{10,1}$.
\item
  Step 4 was omitted (except to set $a_k$ equal to $A_{k1}$), because we had already generated $\bar{A}_{10,1}$.
\item
  In step 6, we did not generate $W_{k1}$, because we had already generated it (see above).
  Note that this ensures that when $H_{k1}$ is fixed, the only randomness in $U_{H_k^{\bar{a}_k}, 1}$ comes from the $H_{kj}$ values ($j=2, \ldots, m$) of the matches.
\item
  In step 8, we set $Z_{Y_{k+1}^{\bar{a}_k}, 1} = \rho_k Z_{H_k^{\bar{a}_k}, 1} + S_{k1}$.
  Again, note that this ensures that when $H_{k1}$ is fixed, the only randomness in $Z_{Y_{k+1}^{\bar{a}_k}, 1}$ comes from the $H_{kj}$ values ($j=2, \ldots, m$) of the matches.
\end{itemize}

Using each of the five $m$ values (Table~\ref{tab:m.values}) in turn, we applied the extended algorithm with these modifications.
We then repeated this procedure using a different random number seed.

We shall refer to the scenarios where $\beta_{k0} = -4.1$ (and so 90\% of individuals survive), $\beta_{k0} = -2.5$ (50\% survive) and $\beta_{k0} = -1.2$ (10\% survive) as the `low-risk', `medium-risk' and `high-risk' scenarios, respectively.
Table~\ref{tab:m.sensitivity} shows, for each scenario, the percentage of the $n=100000$ failure times that were unchanged when we used a different value of $m$ or when we used the same value of $m$ but a different random number seed.
\begin{table}
  \begin{tabular}{r|rrr|rrr}
    & \multicolumn{3}{|c|} {versus $m=4999$} & \multicolumn{3}{c} {versus same $m$} \\
    $m$ & low & med\ & high & low & med\ & high \\ \hline
10  & 86.1  & 63.0  & 10.9  & 90.9  & 64.3 & 100.0 \\
100  & 93.3  & 85.3  & 83.7  & 92.6  & 81.0  & 78.2 \\
500  & 96.7  & 92.7  & 92.0  & 95.9  & 90.6  & 89.4 \\
1000  & 97.5  & 94.6  & 94.0  & 96.9  & 93.1  & 92.3 \\
5000  & 100.0 & 100.0 & 100.0  & 98.6  & 96.9  & 96.6
  \end{tabular}
  \caption{For each of the low-risk, medium-risk and high-risk scenarios, the percentage of generated failure times for $n=100000$ individuals that were the same: a) when $m-1$ matches were used as when 4999 matches were used (columns 2--4); and b) when the same number $m-1$ of matches was used but with two different seeds (columns 5--7).}
  \label{tab:m.sensitivity}
\end{table}

When 4999 matches ($m=5000$) were used, between 97\% (in the medium and high-risk scenarios) and 99\% (in the low-risk scenario) of failure times were unchanged when the random number seed was changed.
When 999 matches ($m=1000$) were used, between 94\% (high-risk scenario) and 98\% (low-risk scenario) of failure times were the same as they had been when 4999 matches were used.
When 499 matches were used, between 92\% and 97\% of failure times were the same as they had been when 4999 matches were used.
When 99 matches were used, between 84\% and 93\% were the same.
Obviously, we would never consider using only nine matches in practice, but, nonetheless, it is interesting to observe that when $m=10$, up to 89\% of failure times were different, confirming that $m=10$ is a poor choice.

The results presented above were obtained using $\rho = -0.9$ as the correlation parameter of the Gaussian copula.
When $\rho$ is smaller (i.e.\ closer to zero) than $-0.9$, the dependence of $Z_{Y_{k+1}^{\bar{a}_k}, 1}$ (in Step 8 of the extended algorithm) on the data generated for the matches will be reduced, suggesting that a smaller $m$ could be used in this case.
Table~\ref{tab:m.sensitivity.05} shows results obtained when $\rho = -0.5$.
\begin{table}
  \begin{tabular}{r|rrr|rrr}
    & \multicolumn{3}{|c|} {versus $m=4999$} & \multicolumn{3}{c} {versus same $m$} \\
    $m$ & low & med\ & high & low & med\ & high \\ \hline
10  &  93.0  & 80.6  & 11.8  & 92.6  & 78.4 & 100.0 \\
100  &  97.4  & 92.9  & 91.0  & 96.7  & 90.6  & 87.9 \\
500  & 98.8  & 96.7  & 95.7  & 98.5  & 95.6  & 94.4 \\
1000  & 99.1  & 97.5  & 96.8  & 98.9  & 96.7  & 95.9 \\
5000  & 100.0 & 100.0 & 100.0  & 98.6  & 96.9  & 96.6
  \end{tabular}
  \caption{Results for $\rho = -0.5$.  For each of the low-risk, medium-risk and high-risk scenarios, the percentage of generated failure times for $n=100000$ individuals that were the same: a) when $m-1$ matches were used as when 4999 matches were used (columns 2--4); and b) when the same number $m-1$ of matches was used but with two different seeds (columns 5--7).}
  \label{tab:m.sensitivity.05}
\end{table}
Here we see that when 499 matches were used, over 95\% of failure times were the same as they had been when 4999 matches were used.

On the basis of these results, we recommend using at least $m=1000$ when the marginal risk of failure is moderate/high and $\rho$ is large.
However, if computation time is an issue, $m=500$ may suffice when the marginal risk is low and/or $\rho$ is small.
In our empirical study, the times taken to generate the data for $n=1000000$ (one million) sampled individuals using $m=5000$ were 3.5, 2.9 and 2.1 hours for the low-risk, medium-risk and high-risk scenarios, respectively.
This was using a single processor of a Dell Latitude 5520 laptop with 4.7GHz 4-core Intel i7 processor.
These $n=1000000$ sampled individuals would suffice for, for example, a simulation study involving 1000 data sets each of 1000 individuals.

\section{Variance estimates in simulation study}

\label{appendix:variance}

For the simulation study reported in Section~\ref{sect:simulationstudy}, Table~\ref{tab:sim.var} shows the empirical SE of the IPTW estimators of $\beta_1,\ldots,\beta_5$, and the means (over the 1000 simulations) of the corresponding estimated SEs calculated using the sandwich variance estimator or bootstrap variance estimator.
Note that when $n=250$, the Monte Carlo SEs of the mean estimated SE calculated from the bootstrap variance estimator can be large relative to those of the mean estimated SE calculated from the sandwich variance estimator.
The maximum Monte Carlo SE for the sandwich method is only 0.012, whereas that for the bootstrap method is 0.057.
This reflects very poor performance of the bootstrap variance estimator in a small number of simulated data sets, a performance which causes considerable bias in the SE estimator for some of the parameter estimators.
Despite this occasional poor performance, the coverage of bootstrap CIs is still better than that of sandwich CIs (see Table~\ref{tab:sim.cover}).

\begin{table}
\begin{tabular}{rrrrrrrrrr}
  \hline
  & \multicolumn{3}{c}{$n=1000$} & \multicolumn{3}{c}{$n=500$} & \multicolumn{3}{c}{$n=250$} \\
  \cmidrule(lr){2-4}\cmidrule(lr){5-7}\cmidrule(lr){8-10}
  & Emp & Sand & Boot & Emp & Sand & Boot & Emp & Sand & Boot \\
  \hline
  \multicolumn{4}{l}{$\delta_{\rm low},\rho_{\rm low}$}\\
  $\beta_1$ & 0.084 & 0.082 & 0.083 & 0.125 & 0.115 & 0.117 & 0.166 & 0.160 & 0.168 \\ 
  $\beta_2$ & 0.170 & 0.167 & 0.167 & 0.243 & 0.234 & 0.238 & 0.331 & 0.326 & 0.343 \\ 
  $\beta_3$ & 0.251 & 0.246 & 0.250 & 0.367 & 0.349 & 0.377 & 0.542 & 0.501 & 1.223 \\ 
  $\beta_4$ & 0.158 & 0.152 & 0.154 & 0.224 & 0.210 & 0.216 & 0.309 & 0.298 & 0.320 \\ 
  $\beta_5$ & 0.323 & 0.310 & 0.317 & 0.465 & 0.439 & 0.467 & 0.670 & 0.626 & 1.310 \\ 
  \hline
  \multicolumn{4}{l}{$\delta_{\rm high},\rho_{\rm low}$}\\
  $\beta_1$ & 0.157 & 0.132 & 0.137 & 0.198 & 0.169 & 0.180 & 0.253 & 0.216 & 0.241 \\ 
  $\beta_2$ & 0.319 & 0.277 & 0.288 & 0.417 & 0.353 & 0.379 & 0.555 & 0.459 & 0.533 \\ 
  $\beta_3$ & 0.371 & 0.326 & 0.337 & 0.503 & 0.422 & 0.458 & 0.652 & 0.560 & 0.837 \\ 
  $\beta_4$ & 0.232 & 0.199 & 0.203 & 0.290 & 0.257 & 0.272 & 0.384 & 0.337 & 0.373 \\ 
  $\beta_5$ & 0.468 & 0.414 & 0.428 & 0.641 & 0.539 & 0.581 & 0.854 & 0.710 & 0.969 \\ 
  \hline
  \multicolumn{4}{l}{$\delta_{\rm low},\rho_{\rm high}$}\\
  $\beta_1$ & 0.084 & 0.084 & 0.084 & 0.121 & 0.118 & 0.120 & 0.172 & 0.165 & 0.173 \\ 
  $\beta_2$ & 0.169 & 0.169 & 0.169 & 0.242 & 0.238 & 0.242 & 0.351 & 0.336 & 0.357 \\ 
  $\beta_3$ & 0.232 & 0.225 & 0.224 & 0.316 & 0.320 & 0.328 & 0.486 & 0.462 & 0.786 \\ 
  $\beta_4$ & 0.142 & 0.139 & 0.141 & 0.195 & 0.196 & 0.202 & 0.295 & 0.282 & 0.301 \\ 
  $\beta_5$ & 0.292 & 0.282 & 0.287 & 0.422 & 0.401 & 0.418 & 0.641 & 0.578 & 0.893 \\ 
  \hline
  \multicolumn{4}{l}{$\delta_{\rm high},\rho_{\rm high}$}\\
  $\beta_1$ & 0.166 & 0.148 & 0.151 & 0.208 & 0.187 & 0.199 & 0.281 & 0.240 & 0.275 \\ 
  $\beta_2$ & 0.349 & 0.304 & 0.316 & 0.470 & 0.396 & 0.431 & 0.606 & 0.522 & 0.910 \\ 
  $\beta_3$ & 0.352 & 0.315 & 0.322 & 0.481 & 0.419 & 0.449 & 0.630 & 0.562 & 0.932 \\ 
  $\beta_4$ & 0.211 & 0.192 & 0.196 & 0.269 & 0.247 & 0.261 & 0.380 & 0.329 & 0.368 \\ 
  $\beta_5$ & 0.447 & 0.397 & 0.410 & 0.610 & 0.525 & 0.564 & 0.794 & 0.704 & 1.059 \\ 
    \hline
\end{tabular}
\caption{Simulation study results.  Empirical SE of IPTW estimators of $\beta_1,\ldots,\beta_5$ (`Emp'), and mean of estimated SE calculated using the sandwich variance estimator (`Sand') or bootstrap (`Boot').  The maximum Monte Carlo SE associated with these mean estimated SEs is 0.007 and 0.010 when $n=1000$ and $n=500$, respectively.  When $n=250$, the maximum Monte Carlo SE is 0.012 for the sandwich method and 0.057 for the bootstrap method.}
  \label{tab:sim.var}
\end{table}

\section{Simulating data for a continuous-time MSM}

\label{appendix:continuous}

In this appendix, we present an algorithm for generating data consistent with a MSM for a survival time outcome when the confounders and treatment can change value at different times and the times at which they change are different for different individuals.
This algorithm could be used to simulate data for the continuous-time MSMs considered by Saarela et al.\ (2016)\cite{Saarela2016}, Ryalen et al.\ (2019, 2020)\cite{Ryalen2019,Ryalen2020} and Dong (2021)\cite{Dong2021}, but can also be used to simulate data for more general continuous-time MSMs.
Unlike the data-generation methods used for simulation studies by the forementioned authors, our algorithm allows the MSM of interest and the values of its parameters to be specified directly.
Saarela et al.\ (2016) ensure that their chosen MSM is approximately correctly specified by assuming that the hazard of a change in the confounder does not depend on past or current treatment and that failure is rare.
Ryalen et al.\ (2019) ensure that their chosen MSM is correctly specified by assuming that the hazard of a change in the exposure given the history of the exposure and confounder depends only on the current value of the confounder and, likewise, the hazard of a change in the confounder depends only on the current exposure value.
Dong (2021) similarly makes restrictive assumptions about the data-generating mechanism; it is also not clear why the assumed MSM would be correctly specified for her data-generating mechanism.
Our algorithm allows a much richer choice of data-generating mechanism than those used by Saarela et al.\ (2016), Ryalen et al.\ (2019) and Dong (2021).

Note that, although in principle the algorithm that we now describe places no restriction on the number of times that an individual's treatment and confounder values change over time, this algorithm could be computationally expensive if such changes were frequent.

Let $\tau$ denote an administrative censoring time; we shall not generate data beyond time $\tau$.
Let $X$ denote baseline variables that will be included as covariates in the MSM of interest.
Let $B$ denote baseline variables that are not included as covariates in the MSM.
Just as in Section~\ref{sect:notation}, $B$ can include baseline confounders, common causes of the confounder and failure processes that are not confounders, and instrumental variables.
Denote an individual's treatment and time-dependent confounder processes as $\big( A(t): t \geq 0 \big)$ and $\big( L(t): t \geq 0 \big)$, respectively.
We assume that $A(t)$ is discrete.
Let $\big( N^A(t): t \geq 0 \big)$ and $\big( N^L(t): t \geq 0 \big)$ denote the counting processes that jump when, respectively, $A(t)$ and $L(t)$ change value, i.e.\ $dN^A(t) = I \{ A(t) \neq \lim_{s \rightarrow \tm} A(s) \}$ and $dN^L(t) = I \{ L(t) \neq \lim_{s \rightarrow \tm} L(s) \}$.
Let $T$ denote the individual's failure time, and let $\big( N^T (t): t \geq 0 \big)$ denote the counting process that jumps when the individual fails, i.e.\ $N^T(t) = I(T \leq t)$.
Let $\bar{A}(t) = \big\{ A(s): 0 \leq s \leq t \big\}$, $\bar{L}(t) = \big\{ L(s): 0 \leq s \leq t \big\}$ and $\bar{N}^T (t) = \big\{ N^T(s): 0 \leq s \leq t \big\}$.
Define $\bar{A}(\tm) = \big\{ A(s): 0 \leq s < t \big\}$ and similarly $\bar{L}(\tm)$ and $\bar{N}^T (\tm)$.\footnote{Readers familiar with Ryalen et al.\ (2019, 2020) may notice that Ryalen et al.\ describe the confounder process in a different way from that used in this paragraph.
  In their formulation, rather than having one counting process that describes the times at which the value of the vector of confounders changes, they would have one counting process for each possible value of the (assumed discrete) confounder vector.
  In fact, they concentrate on the specific scenario where there is only one confounder, whose value is initially 0, may change later to 1, and cannot subsequently change back from 1 to 0.}

Let $\mathcal{F}_t = \sigma \{ X, B, \bar{A}(t), \bar{L}(t), \bar{N}^T(t) \}$ denote the sigma algebra generated by $X$, $B$, $\bar{A}(t)$, $\bar{L}(t)$ and $\bar{N}^T(t)$.
So, $\{ \mathcal{F}_t : t \geq 0 \}$ is the filtration generated by $X$, $B$ and the processes $\big( A(t): t \geq 0 \big)$, $\big( L(t): t \geq 0 \big)$ and $\big( N^T (t): t \geq 0 \big)$.

Let $\lambda^A(t \mid \mathcal{F}_{\tm})$ denote the intensity at time $t$ for the counting process $\big( N^A (t) : t \geq 0 \big)$ with respect to filtration $\mathcal{F}_t$, and write this intensity in the form $\lambda^A(t \mid \mathcal{F}_{\tm}) = \{ 1  - N^T(\tm) \} \times \alpha^A \{ t \mid X, B, \bar{A} (\tm), \bar{L} (\tm) \}$.
Let $p^A \{ a(t) \mid X, B, \bar{A} (\tm), \bar{L} (\tm), dN^A(t) = 1 \}$ denote the probability mass function of $A(t)$ given the history and that a change in treatment occurs at time $t$.
Similarly, let $\lambda^L (t \mid \mathcal{F}_{\tm}) =  \{ 1 - N^T(\tm) \} \times \alpha^L \{ t \mid X, B, \bar{A} (\tm), \bar{L} (\tm) \}$ denote the intensity at time $t$ for $\big( N^L (t) : t \geq 0 \big)$ with respect to $\mathcal{F}_t$, and let $p^L \{ l(t) \mid X, B, \bar{A} (\tm), \bar{L} (\tm), dN^L(t) = 1 \}$
denote the probability mass/density function of $L(t)$ given the history and that a change in confounders occurs at time $t$.
Let $\lambda^T (t \mid \mathcal{F}_{\tm}) = \{ 1 - N^T(\tm) \} \times \alpha^T \{ t \mid X, B, \bar{A} (\tm), \bar{L} (\tm) \}$ denote the intensity at time $t$ for the process $\big( N^T (t) : t \geq 0 \big)$ with respect to $\mathcal{F}_t$.
So, $\alpha^T \{ t \mid X, B, \bar{A} (\tm), \bar{L} (\tm) \}$ is the hazard of failure at time $t$ given $X$, $B$ and the exposure and confounder histories.

Let $\mathcal{F}^*_t = \sigma \{ X, \bar{A}(t), \bar{N}^T(t) \}$ denote the sigma algebra generated by $X$, $\bar{A}(t)$ and $\bar{N}^T(t)$.
So, $\{ \mathcal{F}^*_t : t \geq 0 \}$ is the filtration generated by $X$ and processes $\big( A(t): t \geq 0 \big)$ and $\big( N^T (t): t \geq 0 \big)$.
Clearly, $\mathcal{F}^*_t$ is a sub-sigma algebra of $\mathcal{F}_t$.
Let $\lambda^{*T} (t \mid \mathcal{F}^*_{\tm}) = \{ 1 - N^T (\tm) \} \times \alpha^{*T} \{ t \mid X, \bar{A}(\tm) \}$ denote the intensity of the process $\big( N^T (t): t \geq 0 \big)$ with respect to filtration $\mathcal{F}^*_t$.
So, $\alpha^{*T} \{ t \mid X, \bar{A}(\tm) \}$ is the hazard of failure given $X$ and the exposure history.

Analogously to Ryalen et al.\ (2019) and Dong (2021, page 30), we make the local independence assumption that intervening to set $\bar{A} (\tau)$ equal to some value $\bar{a} = \bar{a} (\tau)$ does not change the intensity function $\lambda^L (t \mid \mathcal{F}_{\tm})$, the probability mass/density function $p^L \{ l(t) \mid X, B, \bar{A} (\tm), \bar{L} (\tm), dN^L(t) = 1 \}$ or the intensity function $\lambda^T (t \mid \mathcal{F}_{\tm})$.
Hence, in particular,
$
  \alpha^T \{ t \mid X=x, B=b, \bar{A} (\tm) = \bar{a} (\tm), \bar{L} (\tm) = \bar{l} (\tm) \}
$
is equal to
$
\alpha^{T, \bar{a}} \{ t \mid X=x, B=b,, \bar{L} (\tm) = \bar{l} (\tm) \},
$
the hazard of failure at time $t$ given $X=x$, $B=b$ and $\bar{L} (\tm) = \bar{l} (\tm)$ when we intervene to set $\bar{A} (\tau) = \bar{a}$.

Let $\lambda^{*T, \bar{a}} (t \mid \mathcal{F}^*_{\tm}) = \{ 1 - N^T (\tm) \} \times \alpha^{*T, \bar{a}} (t \mid X)$ denote the intensity of the process $\big( N^T(t): t \geq 0 \big)$ with respect to filtration $\mathcal{F}^*_t$ when we intervene to set $\bar{A} (\tau) = \bar{a}$.

The user of our algorithm needs to specify:
\begin{itemize}
\item
  the intensity $\alpha^A \{ t \mid X, B, \bar{A} (\tm), \bar{L} (\tm) \}$ and the probability mass function $p^A \{ a(t) \mid X, B, \bar{A} (\tm), \bar{L} (\tm), dN^A(t) = 1 \}$;
\item
  the intensity $\alpha^L \{ t \mid X, B, \bar{A} (\tm), \bar{L} (\tm) \}$ and the probability mass/density function $p^L \{ l(t) \mid X, B, \bar{A} (\tm), \bar{L} (\tm), dN^L(t) = 1 \}$;
\item
  a `risk score' function $h^{\bar{a}} \{ t \mid x, b, \bar{l} (\tm) \}$ of $x$, $b$, $\bar{l} (\tm)$ and $\bar{a} (\tm)$, which will order individuals with the same value of $X$ but different values of $(B, \bar{L}(\tm))$ according to their potential hazards $\alpha^{T, \bar{a}} \{ t \mid X, B,, \bar{L} (\tm) \}$.\footnote{As in Section~\ref{sect:K1}, $h^{\bar{a}} \{ t \mid x, b, \bar{l} (\tm) \}$ can be replaced by $\nu_t [ h^{\bar{a}} \{ t \mid x, b, \bar{l} (\tm) \} ]$, where $\nu_t$ is any monotonically increasing function, because only the ranking of $(B, \bar{L}(\tm))$ matters.}
\item
  the correlation parameter $\rho$ ($-1 \leq \rho \leq 0$) of a Gaussian copula for the association between the risk score and the failure time;
\item
  the hazard $\alpha^{*T, \bar{a}} (t \mid X)$.
\end{itemize}
The last of these would be chosen so that the MSM of interest is correctly specified.
Dong (2021, page 30) gives the example of $\alpha^{T, \bar{a}} (t \mid X) = \exp \{ \theta_0 (t) + \theta_1 a(\tm) \}$ for some $\theta_1$ and function $\theta_0 (t)$ of $t$.
Ryalen et al.\ (2019, equation (1)) assume $\alpha^{T, \bar{a}} (t \mid X) = b \{ X, a(\tm) \}^\top \beta (t)$ for some function $b \{ X, a(\tm) \}$ of baseline covariates $X$ and current treatment $A(\tm)$, and some function $\beta (t)$ of $t$.
In their simulation study (Section 4.5), $\beta(t) = \beta$ is a constant.

Let $T^{\bar{a}}$ denote the failure time when we intervene to set $\bar{A} (\tau) =\bar{a}$.
For $s < t \leq \tau$, define
\[
F_{T^{\bar{a}}} (t \mid X=x, T^{\bar{a}} > s)
=
P (T^{\bar{a}} \leq t \mid X=x, T^{\bar{a}} > s)
=
1 - \int_s^t \exp \{ - \alpha^{* T, \bar{a}} ( u \mid x ) \} \; du.
\]
In order for this function $F_{T^{\bar{a}}} (t \mid X=x, T^{\bar{a}} > s)$ to be a (conditional) CDF, we need to define its value also for $t > \tau$.
Because we administratively censor failure times at time $\tau$, it will not be important exactly how we define $F_{T^{\bar{a}}} (t \mid X=x, T^{\bar{a}} > s)$ for $t>\tau$, provided that it is a continuous increasing function with $\lim_{t \rightarrow \infty} F_{T^{\bar{a}}} (t \mid X=x, T^{\bar{a}} > s) = 1$.
We can achieve this by, for example, defining $F_{T^{\bar{a}}} (t \mid X=x, T^{\bar{a}} > s)$ for $t > \tau$ as the minimum of $F_{T^{\bar{a}}} (\tau \mid X=x, T^{\bar{a}} > s) + t - \tau$ and 1.

For any fixed $x$ and $s$, denote the inverse of $F_{T^{\bar{a}}} (t \mid X=x, T^{\bar{a}} > s)$ as $F_{T^{\bar{a}}}^{-1} (u \mid X=x, T^{\bar{a}} > s)$.\footnote{That is, $F_{T^{\bar{a}}}^{-1} (u \mid X=x, T^{\bar{a}} > s)$ satisfies $F_{T^{\bar{a}}}^{-1} \{ F_{T^{\bar{a}}} (t \mid X=x, T^{\bar{a}} > s) \mid X=x, T^{\bar{a}} > s \} = t$.}

Let $H^{\bar{a}} (t) = \lim_{\delta \rightarrow \zerop} h^{\bar{a}} [ t + \delta \mid X, B, \bar{L} \{ \tdeltam \} ]$.\footnote{This random variable is a function of $t$, $X$, $B$ and $\bar{L} (t)$}
Let $F_{H^{\bar{a}} (t)} (h \mid X=x, T^{\bar{a}} > t) = P \{ H^{\bar{a}} (t) \leq h \mid X=x, T^{\bar{a}} > t \}$ denote the CDF of $H^{\bar{a}} (t)$ given $X=x$ and $T^{\bar{a}} > t$.

Our algorithm assumes that the user's choice of risk score function $h^{\bar{a}} \{ t \mid x, b, \bar{l} (\tm) \}$ satisfies the following condition:
\begin{eqnarray*}
  &&  
\hspace{.5cm} \mbox{If (i) }     
     h^{\bar{a}} \{ t \mid x, b_1, \bar{l}_1 (\tm) \} > h^{\bar{a}} \{ t \mid x, b_2, \bar{l}_2 (\tm) \}
  \nonumber \\
  &&  
     \mbox{and (ii) }
     \big( a(u), l_1(u), l_2(u) \big) = \big( a(\tm), l_1(\tm), l_2(\tm) \big)
     \mbox{ for all } u \mbox{ such that } t \leq u < s,
  \nonumber \\
  &&  
     \mbox{then }
     h^{\bar{a}} \{ s \mid x, b_1, \bar{l}_1 (\sm) \} > h^{\bar{a}} \{ s \mid x, b_2, \bar{l}_2 (\sm) \}.
\end{eqnarray*}

This condition ensures that if one individual has a higher $H^{\bar{a}} (t)$ value than a second individual with the same value of $X$, then the first individual continues to have the higher value of $H^{\bar{a}} (t)$ at all later times until one of these two individuals' time-dependent confounder values changes or $a(t)$ changes.
The condition would be satisfied if, for example,
\begin{equation}
  h^{\bar{a}} \{ t \mid x, b, \bar{l} (\tm) \} =
  \lim_{\delta \rightarrow \zerop} h^{\bar{a}} \big( c(t) + \delta \; \big| \; x, b, \bar{l} [ \{ c(t) + \delta \} \text{-} ] \big)
\label{eq:last.change}
\end{equation}
where $c(t)$ denotes the time of the last change in $\bar{l} (\tm)$, i.e.\ $c(t) = \mbox{argmax}_{s \leq t} \{ dN^L(s) = 1 \}$, with $c(t)=0$ if there is no change.
Equation~(\ref{eq:last.change}) can be interpreted as meaning that the only time that the risk score function $h^{\bar{a}} \{ t \mid x, b, \bar{l} (\tm) \}$ (for a fixed $\bar{a}$) can change is immediately after there is a change in the value of the confounders.

If the CDF $F_{H^{\bar{a}} (t)} (u \mid X=x, T^{\bar{a}} > t)$ were known,\footnote{For example, this CDF is known for the data-generating mechanism used by Ryalen et al.\ (2019) for a simulation study in their Section 4.
  In that study: i) there are no covariates $X$ or $B$; ii) $L(t)$ is binary with $L(0) = 0$ and $L(s)=1 \Rightarrow L(t)=1$ $\forall t>s$; iii) $\alpha^T \{ t \mid \bar{A}(\tm), \bar{L}(\tm) \} = \alpha^T \{ t \mid A(\tm), L(\tm) \}$; and $\alpha^L \{ t \mid \bar{A}(\tm), \bar{L}(\tm) \} = \alpha^L \{ t \mid A(\tm) \}$.
  So, $H^{\bar{a}} (t) = L^{\bar{a}} (\tm)$ is binary, $F_{H^{\bar{a}} (t)} (0 \mid T^{\bar{a}} > t) = P \{ L^{\bar{a}} (t) = 0 \mid T^{\bar{a}} > t \} = \int_0^t \exp [ - \alpha^L \{ s \mid a(\sm) \} ] \; ds$, and $F_{H^{\bar{a}} (t)} (1 \mid T^{\bar{a}} > t) = 1$.}
our algorithm would be as follows.
\begin{enumerate}
\item
  Sample $X$ from $p(X)$.
\item
  Sample $B$ from $p(B \mid X)$.
\item
  Set $T = \infty$ and sample processes $\big( A(t): 0 \leq t \leq \tau \big)$ and $\big( L(t): 0 \leq t \leq \tau \big)$ using intensities $\alpha^A \{ t \mid X, B, \bar{A} (\tm), \bar{L} (\tm) \}$ and $\alpha^L \{ t \mid X, B, \bar{A} (\tm), \bar{L} (\tm) \}$, and probability mass/density functions $p^A \{ a(t) \mid X, B, \bar{A} (\tm), \bar{L} (\tm), dN^A(t) = 1 \}$ and $p^L \{ l(t) \mid X, B, \bar{A} (\tm), \bar{L} (\tm), dN^L(t) = 1 \}$.
  An example to illustrate how this could be done is given immediately after the end of this algorithm.
  Denote the sampled value of $\bar{A} (\tau)$ as $\bar{a}$.
\item
  Set $t=0$.
\item
  Let $b = \mbox{argmin}_{ \{ s: s > t \} } \{ dN^L(s) = 1 \mbox{ or } dN^A(s) = 1 \mbox{ or } s=\tau \}$.
\item
  Calculate $H^{\bar{a}} (t) = \lim_{\delta \rightarrow \zerop} h^{\bar{a}} [ t + \delta \mid X, B, \bar{L} \{ \tdeltam \} ]$.
\item
  Calculate $U_H = F_{H^{\bar{a}} (t)} \{ H^{\bar{a}} (t)  \mid X, T^{\bar{a}} > t \}$ and then $Z_H = \Phi^{-1} (U_H)$.
  Sample $Z_T \sim \mbox{Normal} (\rho Z_H, 1 - \rho^2)$ and calculate $U_T = \Phi ( Z_T )$.
\item
  Calculate $T = t + F_{T^{\bar{a}}}^{-1} (U_T \mid X, T^{\bar{a}} > t)$.
\item
  If $T > b$ and $b < \tau$, set $t=b$ and return to step 5.
\item
  If $T > \tau$, then $T$ is administratively censored at time $\tau$.
  Otherwise, if $T \geq \tau$, then $T$ is the failure time and set $A(t)$ and $L(t)$ equal to `missing' for all $t \geq T$.
\end{enumerate}

In this paragraph, we present a very simple example to illustrate how step 3 could be carried out.
Suppose $A$ is binary and $L$ is a single categorical variable with three levels (0, 1 and 2).
We shall omit $X$.
In order to allow $A(0)$, the initial value of treatment, to depend on $L(0)$, the intial value of the time-dependent confounders, we shall include $L(0)$ in the vector of baseline variables $B$.
For simplicity, suppose $B = L(0)$ is the only baseline variable.
We need to choose a probability mass function for $B$ and a probability mass function $p^A \{ a(0) \mid B \}$ for $A(0)$ given $B$, and sample $B$ and $A(0)$ from these.
Now, suppose that for $t>0$, we choose $\alpha^A \{ t \mid B, \bar{A} (\tm), \bar{L} (\tm) \} = 0.1 + 0.05 L(\tm)$ and $\alpha^L \{ t \mid B, \bar{A} (\tm), \bar{L} (\tm) \} = 0.2 - 0.1 A(\tm)$ and $p^A \{ a(t) \mid B, \bar{A} (\tm), \bar{L} (\tm), dN^A(t) = 1 \} = I \{ a(t) \neq A(\tm) \}$ and $p^L \{ l(t) \mid B, \bar{A} (\tm), \bar{L} (\tm), dN^L(t) = 1 \} = 0.5 I \{ l(t) \neq L(\tm) \}$.
Sample a candidate time to treatment change and a candidate time to confounder change from the $\mbox{exponential} (0.1 + 0.05 L(0))$ and $\mbox{exponential} (0.2 - 0.1 A(0))$ distributions, respectively.
Identify which of these two candidate times is the smaller and call this time $t_1$.
Suppose, for example, that the candidate time to treatment change is the smaller.
Set $A(t_1) = 1 - A(0)$ (note this is equivalent to sampling from probability mass function $I \{ a(t_1) \neq A(t_1 \text{-}) \}$).
Also, set $A(s) = A(0)$ for all $0 < s < t_1$ and set $L(s) = L(0)$ for all $0 < s \leq t_1$.
Now sample candidate time to treatment change and candidate time to confounder change from the $\mbox{exponential} (0.1 + 0.05 L(t_1))$ and $\mbox{exponential} (0.2 - 0.1 A(t_1))$ distributions, respectively.
Identify which of these two candidate times is the smaller and call this time $t_1 + t_2$.
Suppose, for example, that the candidate time to confounder change is the smaller.
Sample $L(t_1 + t_2)$ from probability mass function $0.5 I \{ l(t_1 + t_2) \neq L(t_1) \}$.
Also, set $A(s) = A(t_1)$ for all $t_1 < s \leq t_2$ and set $L(s) = L(t_1)$ for all $t_1 < s < t_2$.
Now sample candidate time to treatment change and candidate time to confounder change from the $\mbox{exponential} (0.1 + 0.05 L(t_1 + t_2))$ and $\mbox{exponential} (0.2 - 0.1 A(t_1 + t_2))$ distributions, respectively.
Identify which of these two candidate times is the smaller and call this time $t_1 + t_2 + t_3$.
Continue in this way until $t_1 + t_2 + \ldots \geq \tau$.

In general, the CDF $F_{H^{\bar{a}} (t)} (u \mid X=x, T^{\bar{a}} > t)$ would be unknown.
Therefore, we propose the following algorithm, which, like the extended algorithm in Section~\ref{sect:manyAX}, uses matches to estimate this CDF at the same time as generating data for a single sampled individual.
To avoid complicating this algorithm, we have chosen not to replace matches that fail with copies of randomly chosen matches that have not yet failed.
This enables the entire treatment and time-dependent confounder processes for the sampled individual (individual $i=1$) and the matches (individuals $i=2, \ldots, m$) to be generated at the very beginning of the algorithm, rather than sequentially during the algorithm, and so reduces the number of steps that we need to write down.
However, the algorithm could be modified to replace failing matches with copies of non-failing matches.
Such modification could be important if the marginal probability of failure before time $\tau$ is high, in which case many of the matches may fail before time $\tau$.

\begin{enumerate}
\item
  Sample $X_1$ from $p(X)$.
  Set $X_j = X_1$ for $j=2, \ldots, m$.
\item
  For $j=1, \ldots, m$, sample $B_j$ from $p(B_j \mid X_j)$.
\item
  Set $T_1 = \infty$ and sample processes $\big( A_1(t): t \geq 0 \big)$ and $\big( L_1(t): t \geq 0 \big)$ using intensities $\alpha^A \{ t \mid X_1, B_1, \bar{A}_1 (\tm), \bar{L}_1 (\tm) \}$ and $\alpha^L \{ t \mid X_1, B_1, \bar{A}_1 (\tm), \bar{L}_1 (\tm) \}$, and probability mass/density functions $p^A \{ a(t) \mid X_1, B_1, \bar{A}_1 (\tm), \bar{L}_1 (\tm), dN^A_1(t) = 1 \}$ and $p^L \{ l(t) \mid X_1, B_1, \bar{A}_1 (\tm), \bar{L}_1 (\tm), dN^L_1(t) = 1 \}$.
  Denote the sampled value of $\bar{A}_1 (\tau)$ as $\bar{a}$.
\item
  For $j=2, \ldots, m$, set $\bar{A}_j (\tau) = \bar{a}$ and $T_j = \infty$, and sample the process $\big( L_j(t): t \geq 0 \big)$ using intensity $\alpha^L \{ t \mid X_j, B_j, \bar{A}_j (\tm), \bar{L}_j (\tm) \}$, and probability mass/density function $p^L \{ l(t) \mid X_j, B_j, \bar{A}_j (\tm), \bar{L}_j (\tm), dN^L_j(t) = 1 \}$.
\item
  Set $t=0$.
\item
  Set $\mathcal{M} = \{ j: \; T_j > t \}$ and let $b = \mbox{argmin}_{ \{ s: s > t \} } \{ dN^L_j(s) = 1 \mbox{ for some } j \in \mathcal{M}, \mbox{ or } dN^A_j(s) = 1 \mbox{ for some } j \in \mathcal{M}, \mbox{ or } s=\tau \}$.
\item
  For $j \in \mathcal{M}$, calculate $H^{\bar{a}}_j (t) = \lim_{\delta \rightarrow \zerop} h^{\bar{a}} [ t + \delta \mid X_j, B_j, \bar{L}_j \{ \tdeltam \} ]$.
\item
  For $j \in \mathcal{M}$, let $R_j$ denote the rank of $H^{\bar{a}}_j (t)$ among the set $\{ H^{\bar{a}}_j (t) : \; j \in \mathcal{M} \}$.
\item
  For $j \in \mathcal{M}$, sample $W_j \sim \mbox{Uniform} (0,1)$, calculate $U_{Hj} = (R_j - W_j) / | \mathcal{M} |$, calculate $Z_{Hj} = \Phi^{-1} (U_{Hj})$, sample $Z_{Tj} \sim \mbox{Normal} (\rho Z_{Hj}, 1 - \rho^2)$, and calculate $U_{Tj} = \Phi ( Z_{Tj} )$.
\item
  For $j \in \mathcal{M}$, calculate $T_j = t + F_{T^{\bar{a}}}^{-1} (U_{Tj} \mid X_j, T^{\bar{a}}_j \geq t)$.
\item
  If $b > \mbox{min} \{ T_j: j \in \mathcal{M} \}$, set $b = \mbox{min} \{ T_j: j \in \mathcal{M} \}$.
\item
  If $T_1 > b$ and $b < \tau$, set $t=b$ and return to step 6.
\item
  If $T_1 > \tau$, then $T_1$ is administratively censored at time $\tau$.
  Otherwise, if $T_1 \leq \tau$, then $T_1$ is the failure time of the sampled individual and set $A_1(t)$ and $L_1(t)$ equal to `missing' for all $t \geq T_1$.
\end{enumerate}

\section{Discrete risk score}

\label{appendix:discrete}

If all the confounders are discrete random variables, then the risk score $H_k^{\bar{a}_k}$, and hence the risk quantile $U_{H_k^{\bar{a}_k}}$, will also be discrete.
The copula requires that $U_{H_k^{\bar{a}_k}}$ be conditionally uniformly distributed on the interval $(0,1)$ given $X$, which cannot be true if $U_{H_k^{\bar{a}_k}}$ is a discrete variable.
To handle this situation, our algorithm can be modified in either of two ways.
We shall assume, without loss of generality, that $H_k^{\bar{a}_k}$ takes integer values.

\subsubsection*{Option 1: Modified Copula}

Instead of setting $U_{H_k^{\bar{a}_k}} = F_{H_k^{\bar{a}_k}} (H_k^{\bar{a}_k} \mid X, Y_k^{\bar{a}_{k-1}}=1)$, generate
\[
  U_{H_k^{\bar{a}_k}} \sim
           \mbox{Uniform} \left(
           F_{H_k^{\bar{a}_k}} (H_k^{\bar{a}_k} - 1 \mid X, Y_k^{\bar{a}_{k-1}}=1), \;
           F_{H_k^{\bar{a}_k}} (H_k^{\bar{a}_k} \mid X, Y_k^{\bar{a}_{k-1}}=1)
           \right).
\]
This ensures that $U_{H_k^{\bar{a}_k}} \mid X \sim \mbox{Uniform}(0,1)$, as required.
If we do this, we are now generating $Y_{k+1}^{\bar{a}_k}$ from
\begin{equation*}
  P(Y_{k+1}^{\bar{a}_k} = 0 \mid X, B, \bar{L}_k) =
  \int_{U_{H_k^{\bar{a}_k}}^{\rm min}}^{U_{H_k^{\bar{a}_k}}^{\rm max}} \;
    \Phi \left(
    \frac{ \Phi^{-1} \{ g_1 (\bar{a}_k, X; \beta) \}  - \rho_k \Phi^{-1} (u) }
    { \sqrt{1 - \rho_k^2} }
    \right) \; du
\end{equation*}
where
\begin{eqnarray*}
  U_{H_k^{\bar{a}_k}}^{\rm min}
  & = &
  F_{H_k^{\bar{a}_k}} (H_k^{\bar{a}_k} - 1 \mid X, Y_k^{\bar{a}_{k-1}}=1)
  \\
  U_{H_k^{\bar{a}_k}}^{\rm max}
  & = &
  F_{H_k^{\bar{a}_k}} (H_k^{\bar{a}_k} \mid X, Y_k^{\bar{a}_{k-1}}=1).
\end{eqnarray*}

\subsubsection*{Option 2: Odds ratio}

When $H_k^{\bar{a}_k}$ is categorical (ordered or unordered) with $J$ levels ($0, \ldots, J-1$), we can dispense with the copula and instead specify the $J-1$ odds ratios:
\[
  \frac{ P(Y_{k+1}^{\bar{a}_k} = 1 \mid X, H_k^{\bar{a}_k} = j) }{ P(Y_{k+1}^{\bar{a}_k} = 0 \mid X, H_k^{\bar{a}_k} = j) }
  \times \frac{ P(Y_{k+1}^{\bar{a}_k} = 0 \mid X, H_k^{\bar{a}_k} = 0) }{ P(Y_{k+1}^{\bar{a}_k} = 1 \mid X, H_k^{\bar{a}_k} = 0) }
  \hspace{0.5cm} (j=1, \ldots, J-1)
\]
The cell probabilities $P(Y_{k+1}^{\bar{a}_k} = y, H_k^{\bar{a}_k} = j \mid X)$ ($y=0,1$; $j=0, \ldots, J-1$) of the $2 \times J$ contingency table for the two categorical variables $Y_{k+1}^{\bar{a}_k}$ and $H_k^{\bar{a}_k}$ can then be calculated from these odds ratios and the marginal probabilities $P(Y_{k+1}^{\bar{a}_k} = y \mid X)$ and $P(H_k^{\bar{a}_k} = j \mid X)$ ($y=0,1$; $j=0, \ldots, J-1$).
If $J=1$ (i.e.\ $H_k^{\bar{a}_k}$ is binary), this is straightforward.
If $J>1$, it could be done using, for example, the Iterative Proportional Fitting algorithm (see \cite{Evans2024} for references).

From these cell probabilities, the conditional probability
\[
  P(Y_{k+1}^{\bar{a}_k} = 0 \mid X, H_k^{\bar{a}_k})
=
  \frac{ P(Y_{k+1}^{\bar{a}_k} = 0, H_k^{\bar{a}_k} \mid X) }
  { P(Y_{k+1}^{\bar{a}_k} = 0, H_k^{\bar{a}_k} \mid X) + P(Y_{k+1}^{\bar{a}_k} = 1, H_k^{\bar{a}_k} = j \mid X) }
\]
can easily be calculated and then $Y_{k+1}^{\bar{a}_k}$ sampled from this probability.

\end{document}